\documentclass{article}
\usepackage{epsfig}
\usepackage{subfigure}
\usepackage{amssymb}
\usepackage{authblk}
\usepackage{multicol}
\usepackage{rotfloat}
\usepackage[breaklinks=true,colorlinks=true,linkcolor=blue]{hyperref}

\usepackage{siunitx}
\usepackage{ragged2e}

\begin{document}
\begin{center}
{\Huge Hall A Annual Report \\ 2014}
\end{center}
\vspace{2cm}
\begin{center}
\includegraphics[width=0.95\textwidth, angle = 0.]{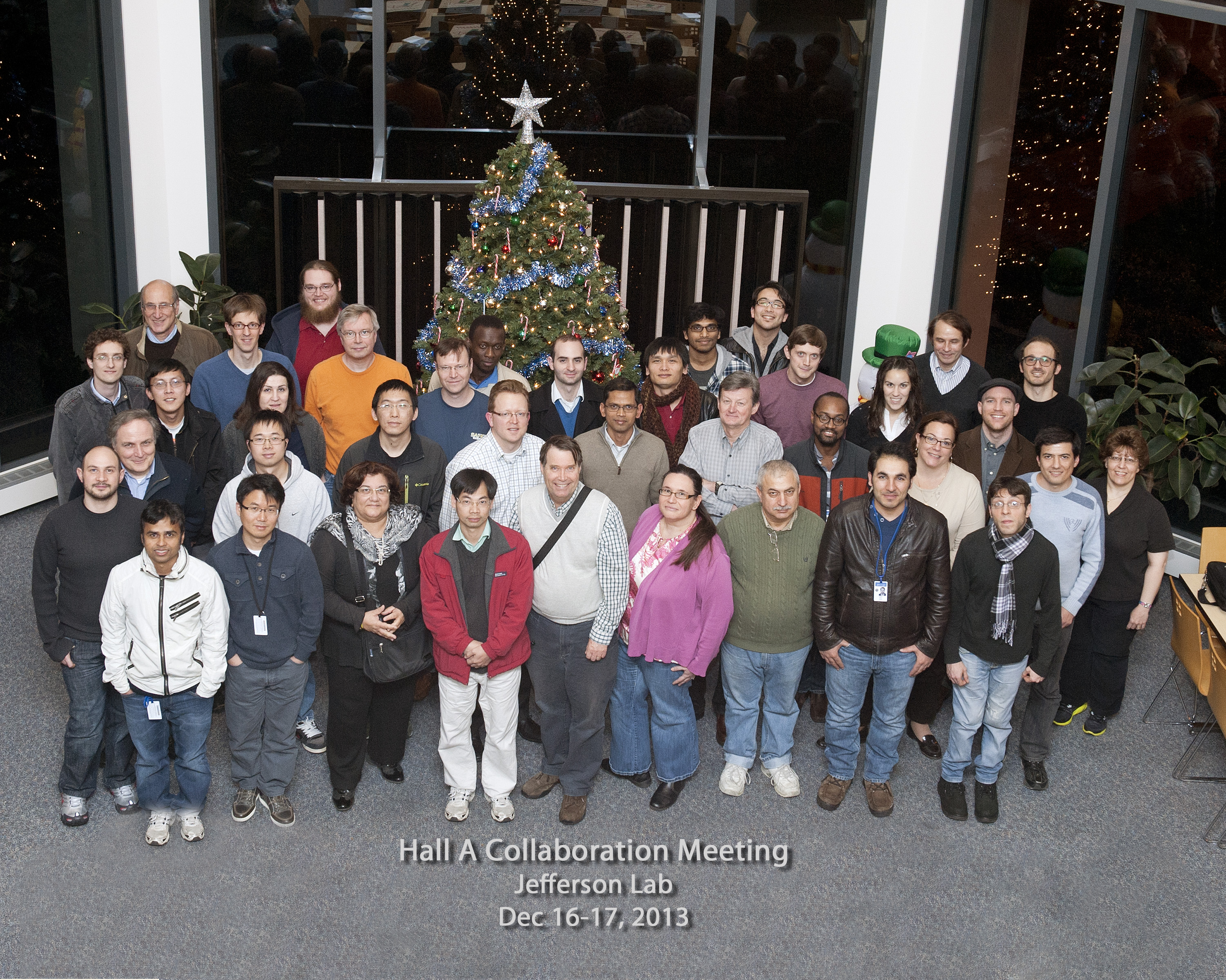}
\end{center}
\vspace{2cm}
\begin{center}
{\Large Edited by Mark M Dalton}
\end{center}
\title{Hall A Annual Report 2014}

\author[1]{M.M.~Dalton} 
\author[1]{E.~Chudakov} 
\author[1]{J.~Gomez}
\author[1]{D. W. Higinbotham}
\author[1]{C.~Keppel}
\author[1]{R.~Michaels}
\author[1]{L.~Myers}
\affil[1]{Thomas Jefferson National Accelerator Facility, Newport News, VA 23606, USA}

\author[2]{K.~Aniol}
\author[2]{S.~Iqbal}
\author[2]{N.~See}
\affil[2]{California State University Los Angeles, Los Angeles, CA 90032, USA}

\author[3]{J.R.~Arrington}
\affil[3]{Argonne National Laboratory, Argonne, IL 60439, USA}

\author[4,5]{M.V.~Ivanov}
\affil[4]{Institute for Nuclear Research and Nuclear Energy, Sofia 1784, Bulgaria}
\affil[5]{Universidad Complutense de Madrid, Madrid E-28040, Spain}

\author[6]{M.~Mihovilovi\v{c}}
\author[6]{S.~\v{S}irca}
\affil[6]{University of Ljubljana, SI-1000 Ljubljana, Slovenia}

\author[7]{N.~Muangma}
\affil[7]{Massachusetts Institute of Technology, Cambridge, MA 02139, USA}

\author[8]{Dien Nguyen}
\affil[8]{University of Virginia, Charlottesville, VA 22901, USA}

\author[9]{R.~Pomatsalyuk}
\author[9]{O.~Glamazdin}
\author[9]{V.~Vereshchaka}
\affil[9]{National Science Center Kharkov Institute of Physics and Technology, Kharkov 61108, Ukraine}

\author[10]{S.~Riordan}
\affil[10]{Stony Brook University, Stony Brook, NY 11794, USA}

\author[11]{T.~Su}
\affil[11]{Kent State University, Kent, OH 44240, USA}

\author[12]{V.~Sulkosky}
\affil[12]{Longwood University, 201 High Street, Farmville, VA 23909, USA}

\author[13]{P.~Zhu}
\affil[13]{University of Science and Technology of China, Hefei 230026, People's Republic of China}

\author[1]{the Jefferson Lab Hall A Collaboration}

\renewcommand\Authands{ and }
\maketitle
\newpage
\tableofcontents
\newpage
\listoffigures
\newpage

\clearpage

\section{Introduction}

\begin{center}
contributed by C. Keppel.
\end{center} 

The year 2014 welcomed the first 12 GeV era beam back to Hall A in two run periods. While there were challenges to address starting up, both for accelerator and for the Hall, all necessary systems were commissioned, and even a bit of physics data was obtained. Hall A Experiments E12-06-114, a measurement of deeply virtual Compton scattering (DVCS), and E12-07-108, a measurement of the proton magnetic form factor G$_M^p$, were the first to receive beam in the 12 GeV era. This would not have been possible without the diligent and expert preparatory work of the Hall A collaboration and staff, and the E12-06-114 and E12-07-108 collaborations. All are to be congratulated!

All detector upgrades to both HRS’s were highly successful, as was the implementation of the DVCS standalone calorimeter and trigger. The hall beamline was largely commissioned, including the Moller polarimeter, the Hall A arc, a new raster system, the revived Unser, as well as the beam position and charge monitors. The Hall suffered from the loss of the HRS-right front quadrupole magnet, which will be repaired or replaced in 2015. The year ended with work still underway on the harps and the Compton polarimeter.

The 12 GeV scientific plans for the hall consist of many compelling experiments to utilize the standard Hall A equipment, some with slight modifications, in conjunction with the higher energy beam. Four (two newly approved this year) require a $^3$H target, one to measure the F$_2^n$/ F$_2^p$ structure function ratio at large x, and one to continue the successful Hall A studies of short range correlation phenomena. This target and associated systems are in design currently for a run after the E12-07-108 and E12-06-114 experiments. Beyond experiments that will utilize the standard Hall A equipment are ambitious plans involving multiple new experiment installations. 

Construction continued this year on one of these larger scale installation experiments, the Super Bigbite Spectrometer (SBS) program. The SBS project consists of a set of three form factor experiments centered around somewhat common equipment and new experimental capabilities, as well as a semi-inclusive experiment focused on quark transverse momentum dependent distributions. The primary SBS spectrometer magnet was delivered from the Brookhaven National Laboratory, modified, and assembled in the Jefferson Lab test lab. Progress was also made on the detector systems, including GEM trackers and hadron and electron calorimeters. A host of scientific development activities for the program are underway, including detector prototype and construction projects, data acquisition upgrades, and refined physics projections. 

Work has continued effectively as well on many other fronts, including infrastructure improvements in data acquisition, offline analysis, and core hall capabilities. Technical preparations are underway for planned experiments such as A$_1^n$, APEX, CREX, PREX-II, and others. In addition to these, three new experiments were approved this year. Two, mentioned above, will round out the $^3$H run group and include a measurement of the electric form factor in the mirror nuclei $^3$He and $^3$H and a measurement of the proton and neutron momentum distributions in these nuclei. The third newly approved experiment is a measurement of the spectral function of $^{40}$Ar through the (e, e $2$p) reaction which is of particular interest to upcoming neutrino experiments. 

Looking farther into future planning, the MOLLER experiment had a highly successful Science Review by the DOE this year. Building on the momentum created by the positive review, the collaboration has further developed the intricate spectrometer design, simulated radiation backgrounds, and progressed plans for the beam line and detector systems. The SOLID experiment submitted a Conceptual Design Report for an anticipated 2015 Director’s Review. This document is substantial, and reflects excellent work by the collaboration, advancing a highly rated experimental program supported by a well-considered, sophisticated detector system. Plans for the future configuration of the Hall to accommodate first the SBS era, followed by MOLLER and SoLID eras, were made. From these, infrastructure work to accommodate these experiments can commence.  

There has been active engagement in analyses of past experiments. Here, fourteen new publications related to Hall A experiments were authored by members of the Hall A collaboration, which included a Physical Review Letter edition with three Hall A articles. In addition, two new Hall A related doctoral theses were successfully defended.

In all, this has been a year of sometimes frustrating, sometimes rewarding, always demanding, work as the first beam was delivered back to the Hall. It continues to be a joy and a privilege to work with the dedicated Hall A community. Please accept my profound thanks to you all for your expert, industrious, innovative efforts. I look forward to continuing the outstanding Hall A program into the 12 GeV era with you!
\clearpage

\section{Hall A Equipment}



\subsection{Hall A BCM System Status }
\label{sec:BCM}

\begin{center}
$^1$R.~Pomatsalyuk, $^1$O.~Glamazdin, $^2$J.~Gomez,  $^2$R.~Michaels, $^2$L.~Myers \\
$^1$National Science Center Kharkov Institute of Physics and Technology, Kharkov 61108, Ukraine  \\
$^2$Thomas Jefferson National Accelerator Facility, Newport News, VA23606, USA  \\
\end{center}

\begin{figure}[!ht]
\begin{center}
\begin{minipage}[t]{0.68\textwidth}
\hrule height 0pt
\includegraphics[width=\textwidth]{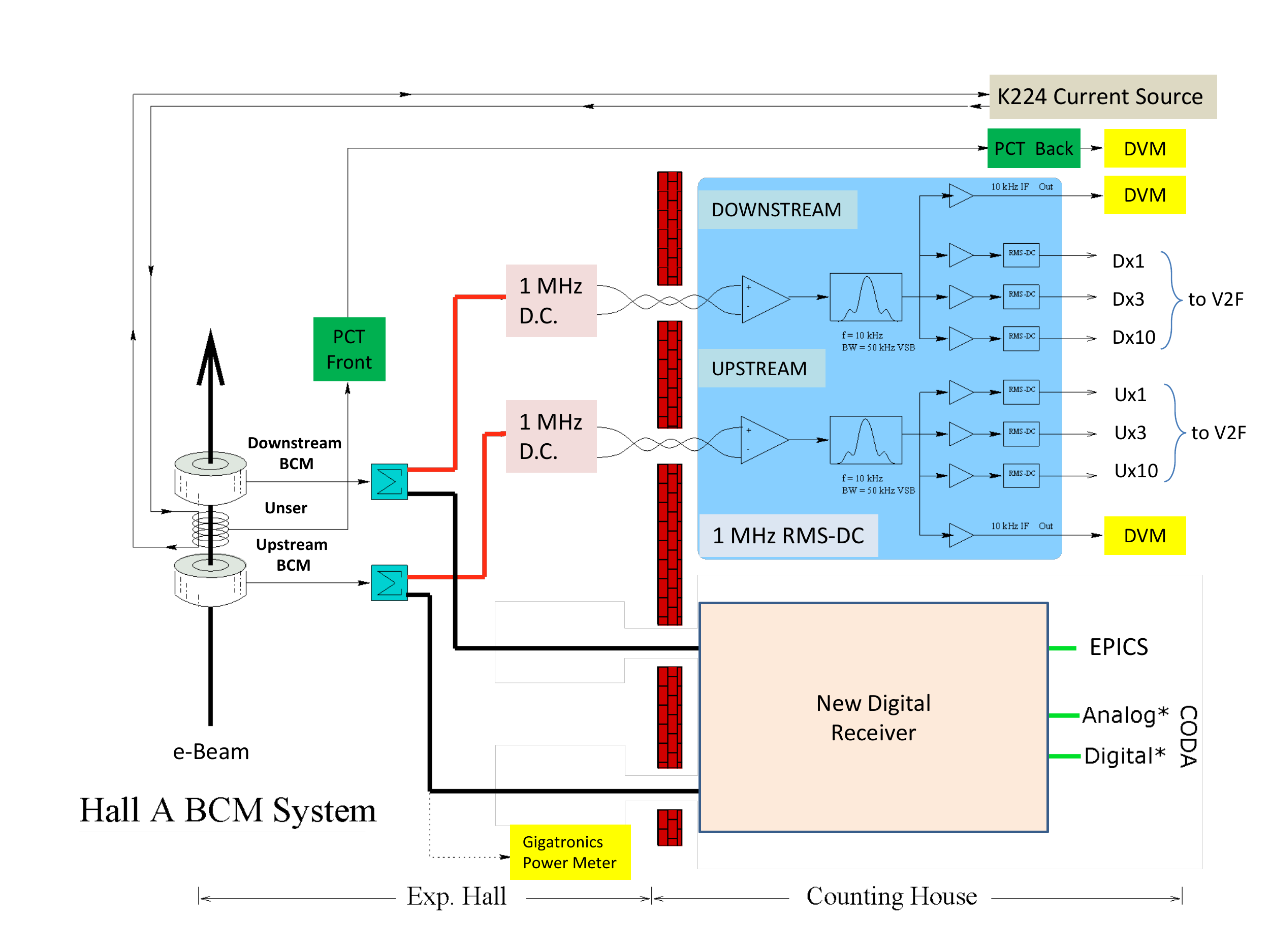}
\end{minipage}
\begin{minipage}[t]{0.3\textwidth}
\hrule height 0pt
\vskip -3mm
  \caption [BCM: The Hall A beam current monitor system.]{The Hall A beam current monitor system.}
   \label{fig:pic1} 
\end{minipage}
\end{center}
\end{figure}

The Hall A beam current monitor system consist of  two cavity monitors and a parametric current transformer (PCT, ''Unser''). It is designed to measure a continues electron beam ($\sim$500~MHz pulse rate). The schematic diagram of Hall A beam current monitor system is shown on Fig.~\ref{fig:pic1}. Two cavity monitors ''Upstream'' and ''Downstream'' are calibrated with respect to a Faraday Cup. The Unser is an absolute measuring device but it suffers of offset drifts equivalent to a couple of $\mu$A ~\cite{denard}. Its calibration can be checked 
 periodically by injecting a known current through a wire crossing the Unser toroid parallel to the beam direction.


In the past a two stage calibration procedure was used:

\begin{itemize}
   \item linearity test of the Hall A BCM cavities with respect to the {\bf Faraday Cup \#2}; 
   \item cross calibration of the Hall A Unser monitor with the {\bf Faraday Cup \#2}. 
\end{itemize}

The cavity monitor signal is split and sent through two parallel paths. One path leads to the old 1~MHz electronic box and another one is connected to a new digital receiver ~\cite{musson}. The AC output of 1~MHz electronic 
blocks is connected to a digital voltmeter (DVM) with EPICS readout. Three voltage-to-frequency converters (VtoF) per cavity provide a value of the instantaneous beam current to the DAQ system. Three VtoF converters are used 
to maximize the dynamic range.  

New digital receiver has FPGA logic on board and digital signal processing (flexibility to be customized for our goals).  It has also analog output from DAC (18~bits, compatible with old system) and digital interface (TTL, optical) for fast data transfer. The new digital receiver can be used also with beam position monitor (BPM). In 
that case it provides measure of an electron beam position.

More information about Hall A BCM/Unser can be found on the web-page.%
\footnote{http://hallaweb.jlab.org/equipment/BCM/}%

\subsubsection{Unser Status}
\label{sec:unserstatus}

\begin{figure}[!ht]
\begin{center}
\begin{minipage}[t]{0.68\textwidth}
\hrule height 0pt
\includegraphics[width=\textwidth]{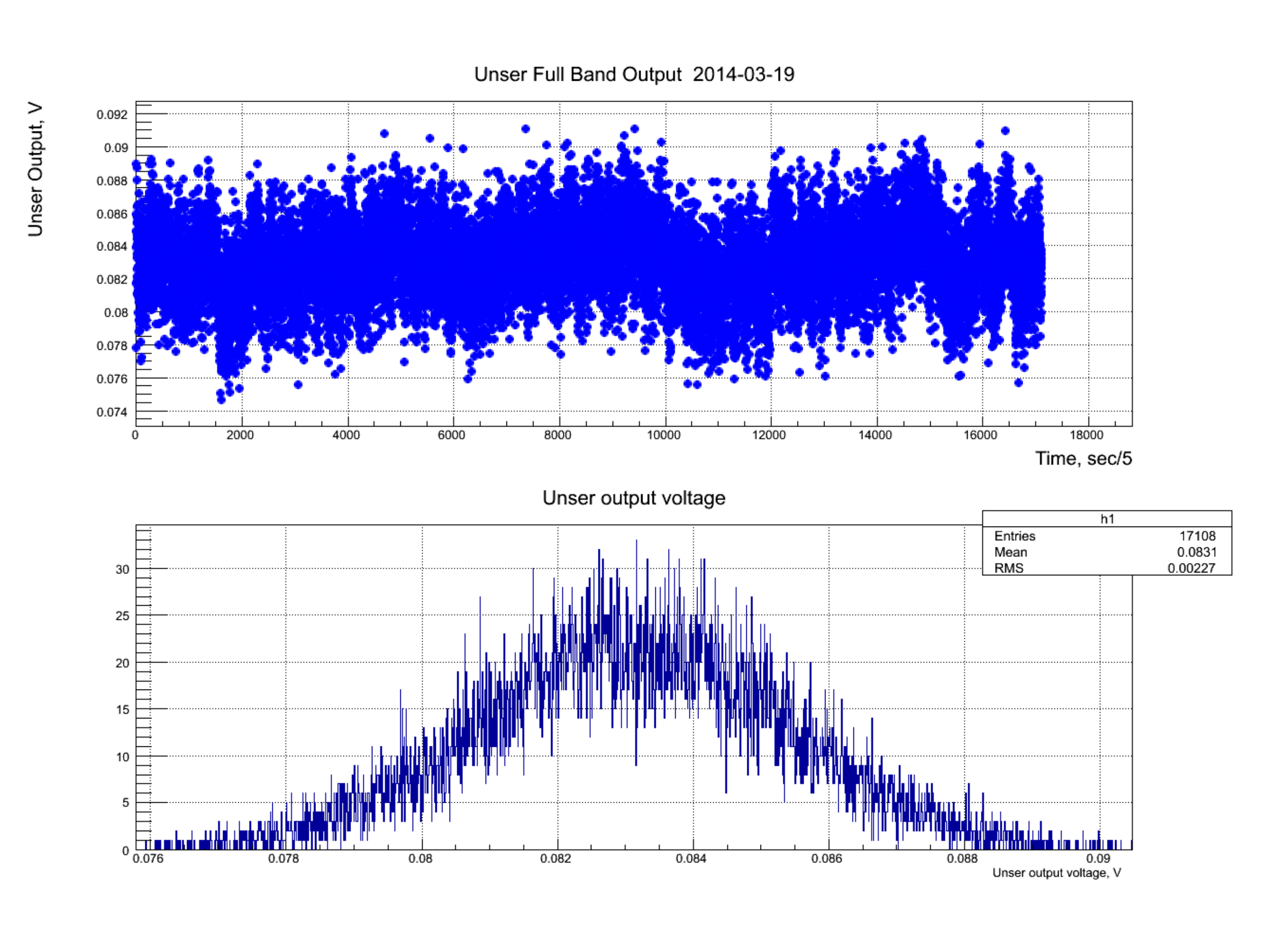}
\end{minipage}
\begin{minipage}[t]{0.3\textwidth}
\hrule height 0pt
\vskip -3mm
   \caption [BCM: Scatter plot and histogram of Unser output offset.]{Scatter plot (top) and histogram (bottom) of Unser output offset measured during 24~hours (with 5~sec time interval).}
   \label{fig:pic2} 
\end{minipage}
\end{center}
\end{figure}

As it was mentioned before, the Unser has offset drifts order of $\pm2~\mu$A rms at constant temperature and a slope of $\pm 5~\mu A / ^\circ$C that cannot be compensated in place. These numbers are not determined by the electronics but by the magnetic material used to build the sensor head (Vitrovac\textregistered6025) ~\cite{unser}. An example of the  Unser drifts taken during a 24~hours period at 5~sec sampling interval is shown on Fig.~\ref{fig:pic2}.

\begin{figure}[!ht]
\begin{center}
\begin{minipage}[t]{0.68\textwidth}
\hrule height 0pt
\includegraphics[width=\textwidth]{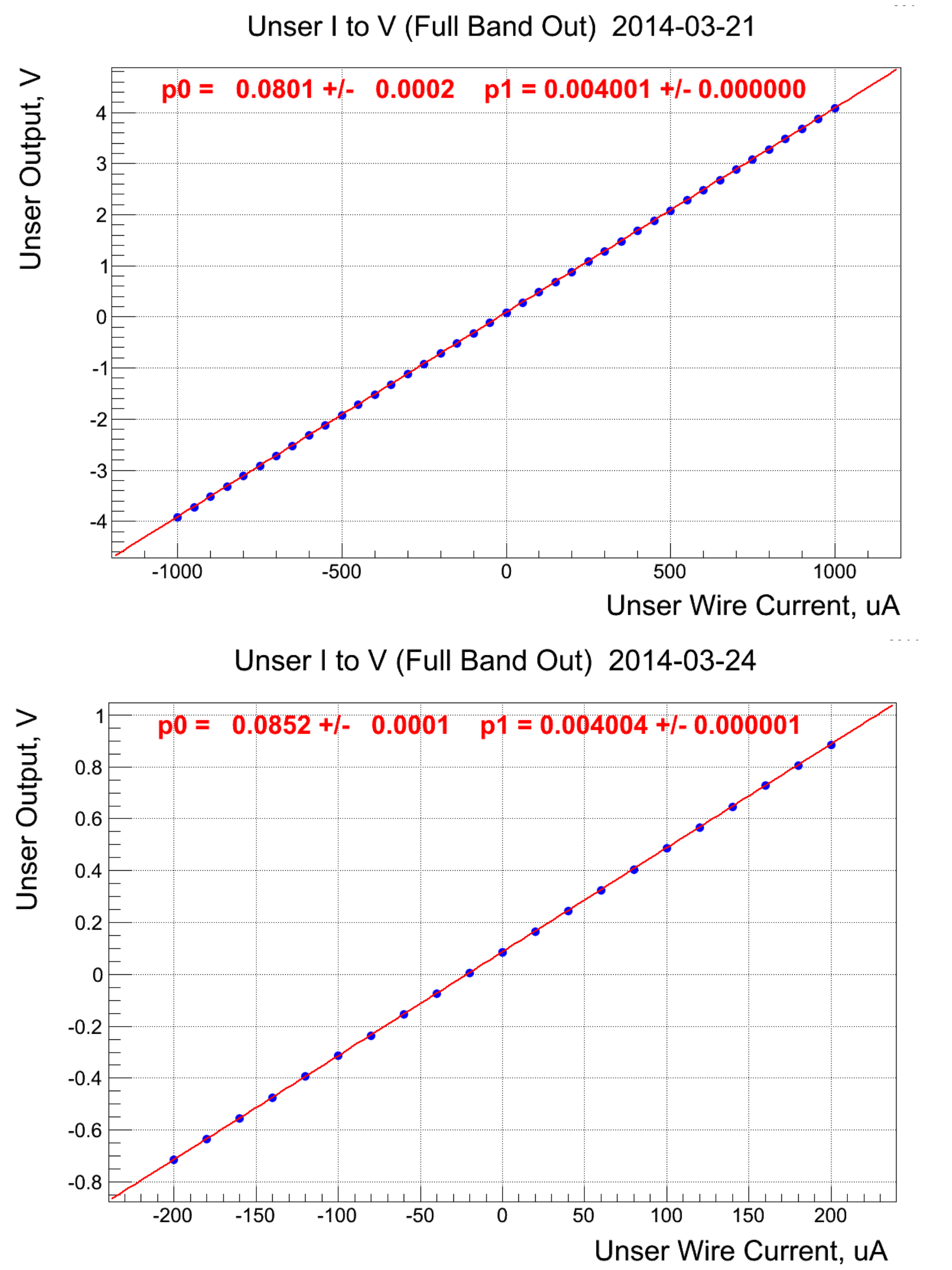}
\end{minipage}
\begin{minipage}[t]{0.3\textwidth}
\hrule height 0pt
\vskip -3mm
   \caption [BCM: Unser wire calibration.]{Unser wire calibration with current source K224 for different ranges.}
   \label{fig:pic3} 
\end{minipage}
\end{center}
\end{figure}



\subsubsection{New BCM receiver test }
\label{sec:receiver}

\begin{figure}[!ht]
\begin{center}
\begin{minipage}[t]{0.68\textwidth}
\hrule height 0pt
\includegraphics[width=\textwidth]{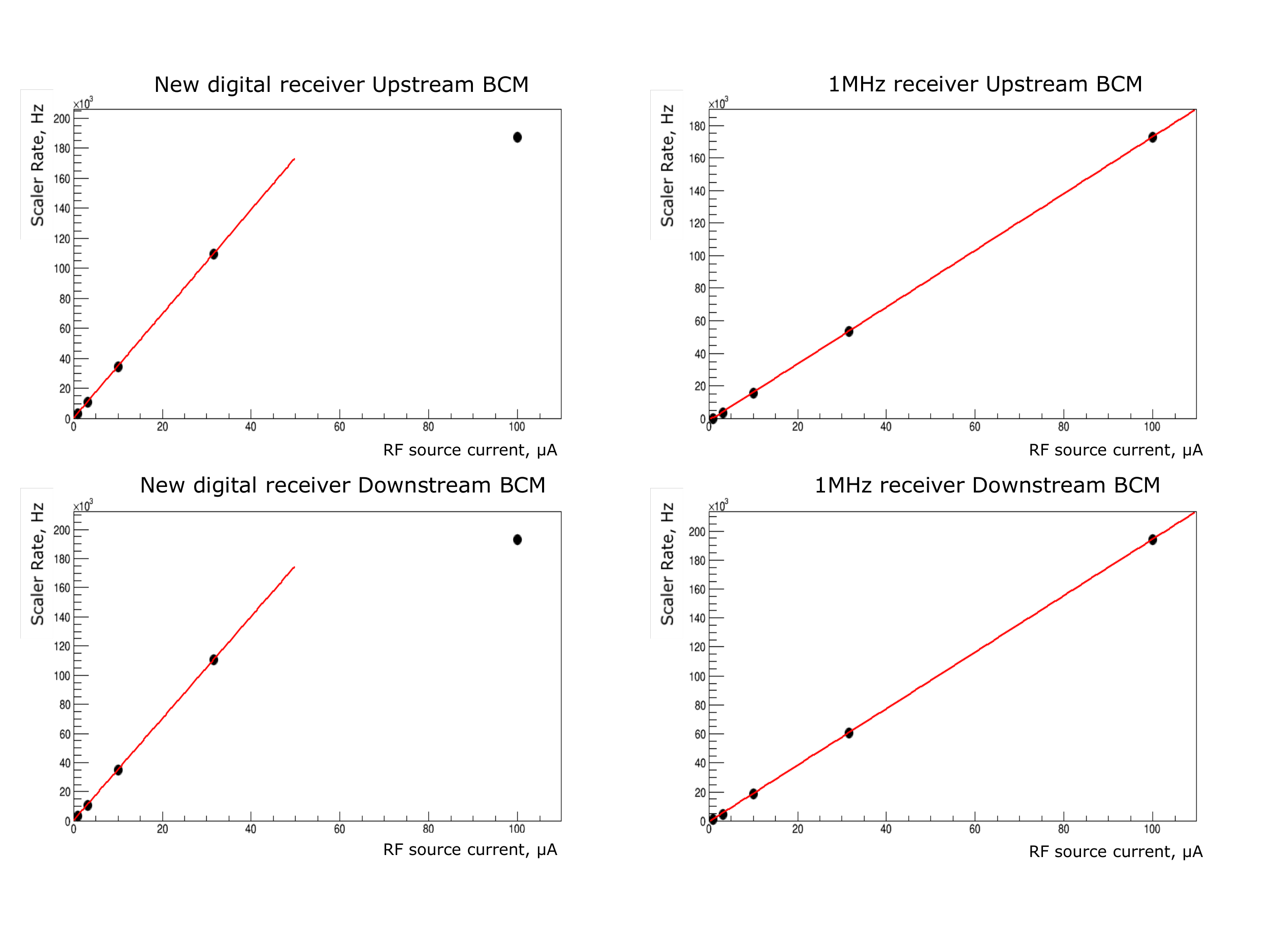}
\end{minipage}
\begin{minipage}[t]{0.3\textwidth}
\hrule height 0pt
\vskip -3mm
   \caption [BCM: New and old BCM receiver calibration.]{The calibration of new digital BCM receiver (left) and old 1~MHz BCM receiver (right) with RF source.}
   \label{fig:pic4} 
\end{minipage}
\end{center}
\end{figure}

New digital BCM receiver was installed on November 2014 in the Hall A. The result of linearity test of both the new digital BCM receiver and the old 1~MHz BCM receivers with RF current source performed by J.~Musson and L.~Myers are shown on Fig.~\ref{fig:pic4}. %
\footnote{HALOG entry \#3306142 } %
Both systems appear to be quite linear (the residuals are less than 1\% above 1~$\mu$A). The new receivers are linear up to 30~$\mu$A. The observed saturation at higher beam current (100~$\mu$A) was due to the chosen electronic gain. The gain can be changed to work at higher current. 


\subsubsection{BCM/BPM Fast Readout for Hall A DAQs (in progress) }
\label{sec:project}

A fast readout of the beam current monitor and position monitor via the new digital receiver system is being develop. Here we summarize the requirement we are working to meet. Two modes of operation are considered,

{\bf Event mode} 

\begin{figure}[!ht]
\begin{center}
\begin{minipage}[t]{0.68\textwidth}
\hrule height 0pt
\includegraphics[width=\textwidth]{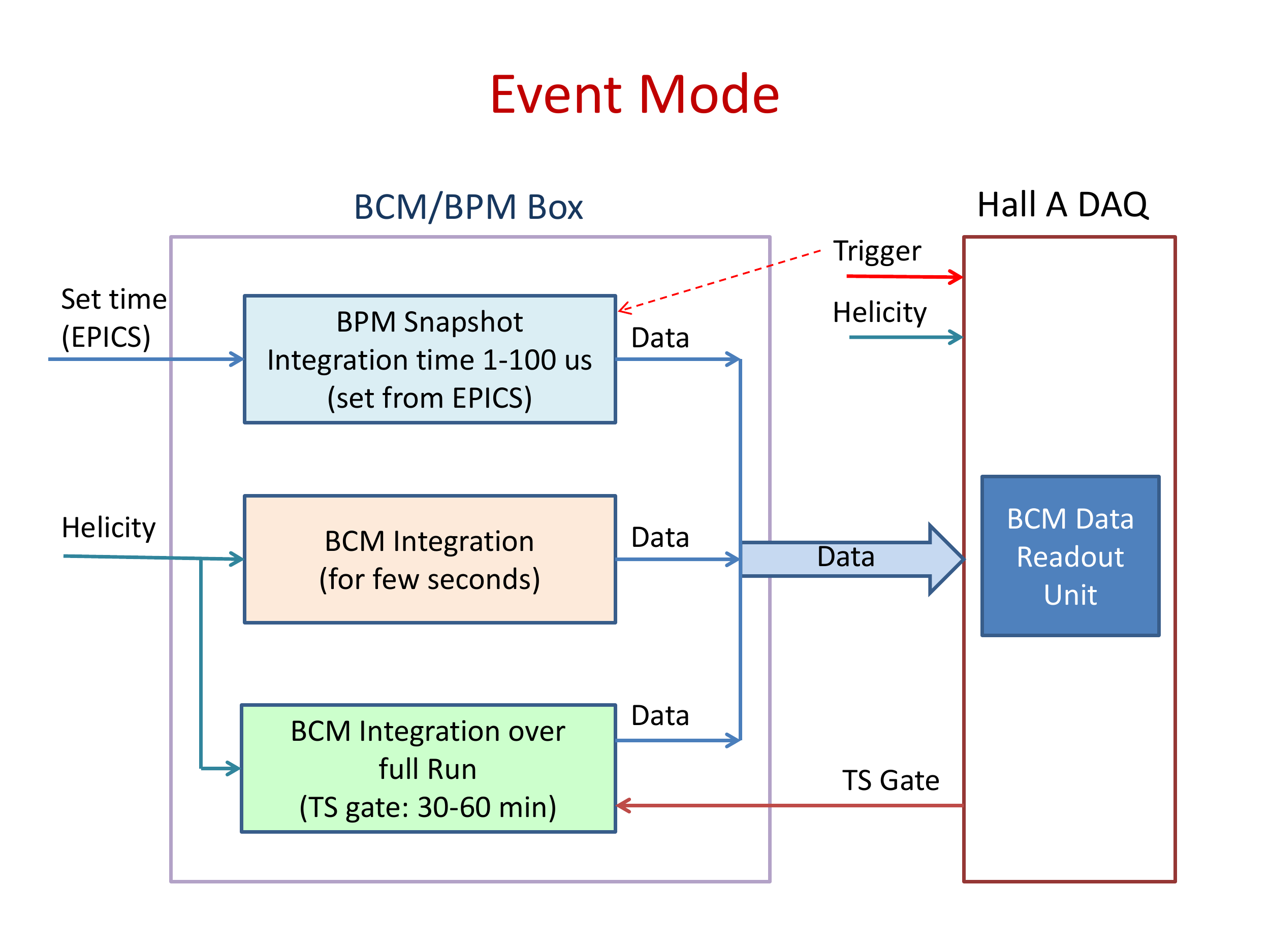}
\end{minipage}
\begin{minipage}[t]{0.3\textwidth}
\hrule height 0pt
\vskip -3mm
   \caption [BCM: Event mode operation diagram.]{Event mode operation diagram.}
   \label{fig:pic5} 
\end{minipage}
\end{center}
\end{figure}

An event is a moment in time defined by a trigger from detectors. Events arrive randomly in time 
(see Fig.~\ref{fig:pic5}).


\begin{itemize}
   \item A snapshot of the BPM data, ideally with 5~$\mu$m resolution (RMS of distribution) and integrated over a time interval $\delta t$ which is smaller than {\it 1/fraster} or {\it 1/fmovement}. Here, {\it fraster} is the raster frequency (25~kHz) and {\it fmovement} is the (presumably low) frequency of other movements of the beam. So, ideally $\delta t \le 4~\mu sec$; if there is a phase lag of a few $\mu$sec between the raster and the data, because this could be calibrated out. 
   \item Other experiments could pick a longer integration time to achieve a higher accuracy per snapshot. It is recommended to let $\delta t$ be a control parameter which would be set infrequently, like maybe at the start of an experiment.
  \item Snapshots of the BCM data with the integration time few seconds.
  \item The beam charge integrated over the time of a ''run''. A run lasts 30 to 60 minutes, typically. The time of a run is provided by a logic gate from the Trigger Supervisor (TS Gate).
\end{itemize}	 

{\bf Parity mode}

\begin{figure}[!ht]
\begin{center}
\begin{minipage}[t]{0.68\textwidth}
\hrule height 0pt
\includegraphics[width=\textwidth]{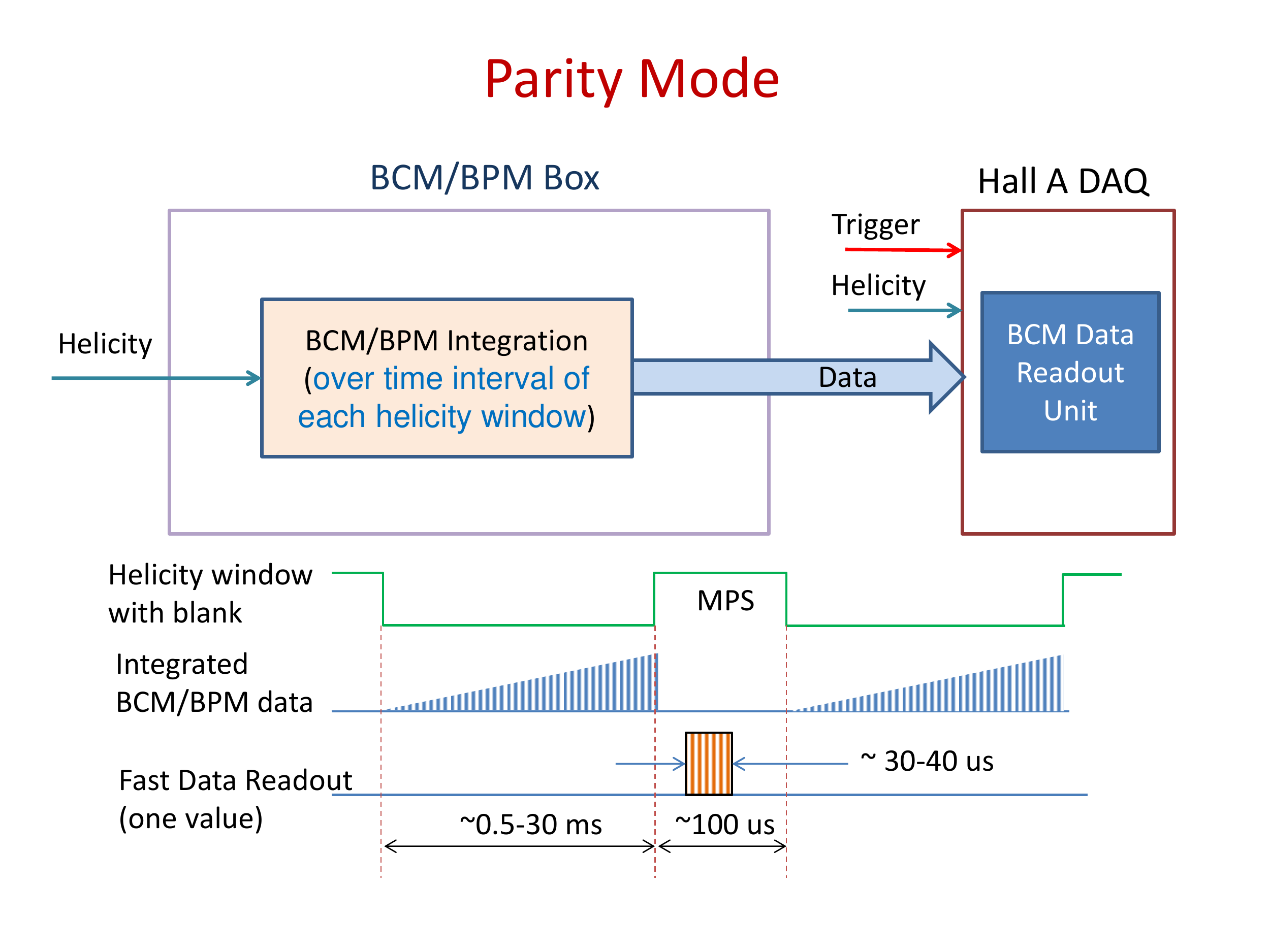}
\end{minipage}
\begin{minipage}[t]{0.3\textwidth}
\hrule height 0pt
\vskip -3mm
   \caption [BCM: Parity mode operation diagram.]{Parity mode operation diagram.}
   \label{fig:pic6} 
\end{minipage}
\end{center}
\end{figure}

The DAQ runs at a fixed frequency governed by the helicity control board. Helicity frequency is in range from 30~Hz up to 2~kHz (see Fig.~\ref{fig:pic6}).

\begin{itemize}
   \item BCM (or BPM) signals are integrated over the interval of the helicity (0.5~-~30~ms). In between   the helicity flips, there is blank off the integration for typically 100~$\mu$sec (MPS) to allow for the Pockels Cell at the polarized source to settle.
   \item The integration gate can be provided by the Hall A DAQ.
\end{itemize}

The consumers of the Event Mode are: 

\begin{itemize}
   \item HRS DAQ;
   \item SBS DAQ;
   \item Third arm DAQs (e.g. Bigbite). 
\end{itemize}

The consumers of the Parity Mode are: 

\begin{itemize}
   \item HAPPEX DAQ;
   \item Moller Polarimeter DAQ;
   \item Compton Polarimeter DAQ. 
\end{itemize}

There could be other DAQ systems that use these data. The BCM data readout unit is supposed to be a custom VME 
unit to receive serial data stream. The unit is under development. 


\clearpage



\subsection{Status of the Hall A M\o{}ller Polarimeter Upgrade }
\label{sec:Moller}

\begin{center}
$^1$O.~Glamazdin, $^2$E.~Chudakov, $^2$J.~Gomez, $^1$R.~Pomatsalyuk, $^1$V.~Vereshchaka \\
$^1$National Science Center Kharkov Institute of Physics and Technology, Kharkov 61108, Ukraine  \\
$^2$Thomas Jefferson National Accelerator Facility, Newport News, VA23606, USA  \\
\end{center}

The Hall A M\o{}ller polarimeter was upgraded in the period
2010~-~2013 in line with the plan to upgrade CEBAF to beam energy of
up to 12~GeV. The main purpose of the upgrade was to expand the
operating energy range of the M\o{}ller polarimeter from
$0.8\div6.0$~GeV before the upgrade to $1.0\div11.0$~GeV after the
upgrade ~\cite{report2012}. The first commissioning run of the
M\o{}ller polarimeter after the upgrade was done in
April 2014 with a beam energy of 6.05~GeV. The test showed that all
elements of the upgraded polarimeter are working properly.

\begin{figure}[hbt]
   \caption [Moller: M\o{}ller events shift on the aperture detector]{
              Left: GEANT-simulated distribution of the M\o{}ller events 
              at the aperture detector.
              Right: amplitude spectra of the signals from the aperture 
              detector measured with the coincidence trigger (both arms 
              aperture and calorimeter detectors in coincidence).}
   \centering
        \includegraphics[width=1.0\textwidth]{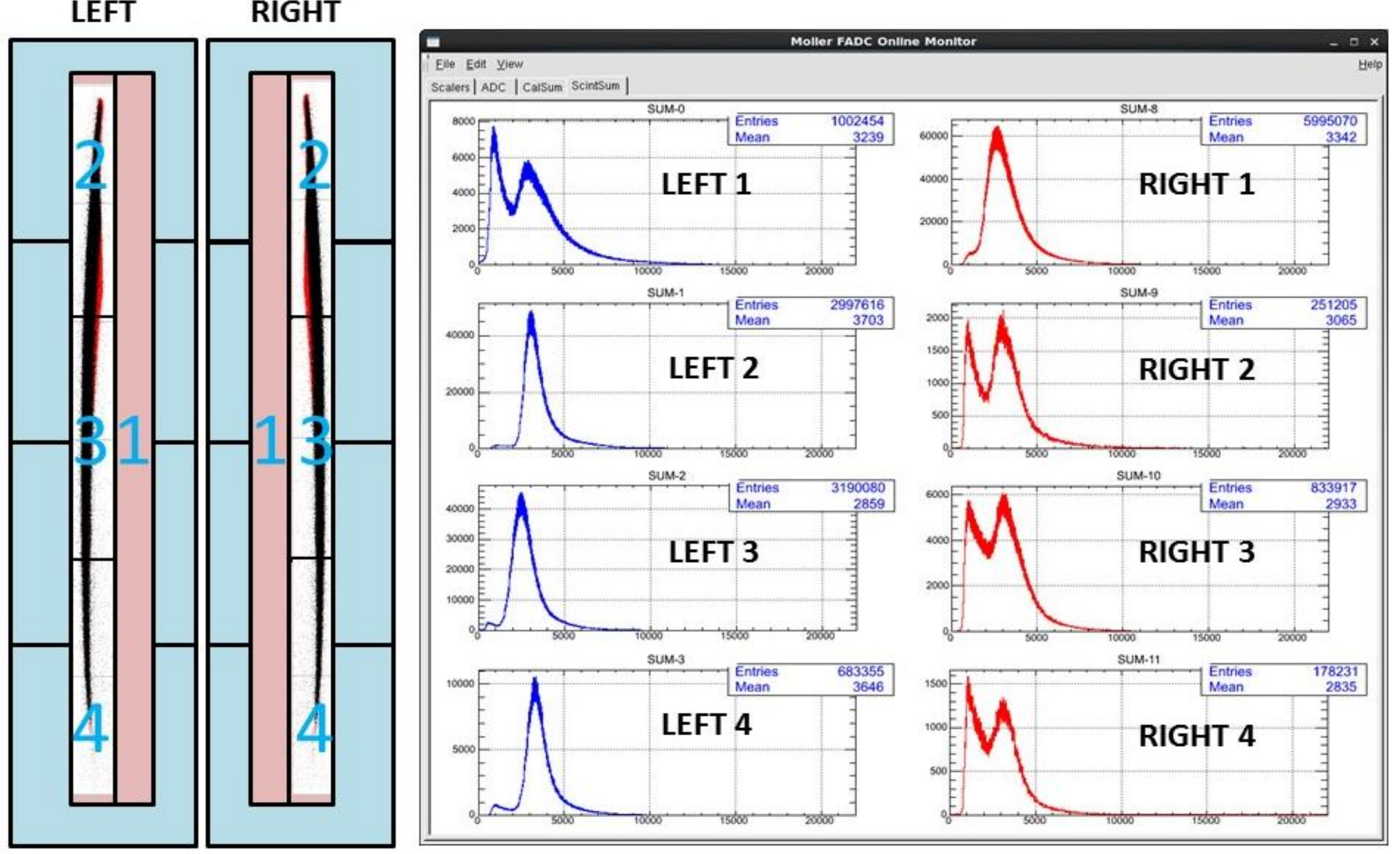}
       \label{fig:detector}
\end{figure}

A significant asymmetry of counting rates between the left and right
arms of the M\o{}ller detector was found while tuning the M\o{}ller
detector. Analysis of the counting rates indicated that total
misalignment of the elements of the polarimeter is up to 10~mm (see
Fig.~\ref{fig:detector}).

Following the commissioning run, the M\o{}ller polarimeter dipole and
detector components positioning were checked by the Jefferson Lab
Survey and Alignment group. The results are listed below,

\begin{itemize}
   \item The lead collimator downstream of the M\o{}ller dipole
     magnet was shifted 4.2~mm to the right of the beam axis;
   \item The slit upstream of the detector was shifted about 1~mm to
     the right;
   \item The rectangular slot upstream of the detector
     shielding box was shifted about 1~mm to the right;
   \item The center of the detector was 2.5~mm to the left of the beam
     axis.
\end{itemize}

Thus, the total measured misalignment of the polarimeter elements was
about 8.7~mm. This value is in good agreement with the prediction
based on the detector counting rates asymmetry. The largest part of
the misalignment - the 4.2~mm shift of the lead collimator
was fixed in summer 2014.

\begin{figure}[hbt]
   \caption [Moller: M\o{}ller dipole shielding.]{TOSCA result for
     the M\o{}ller dipole with the 10~cm extended shielding pipe. A
     picture of the M\o{}ller dipole with additional shielding pipe
     (left picture). TOSCA calculation of the electron beam shift on
     the Hall A target (right picture). The beam energy is shown with
     a red circle.}
   \centering
        \includegraphics[width=1.0\textwidth]{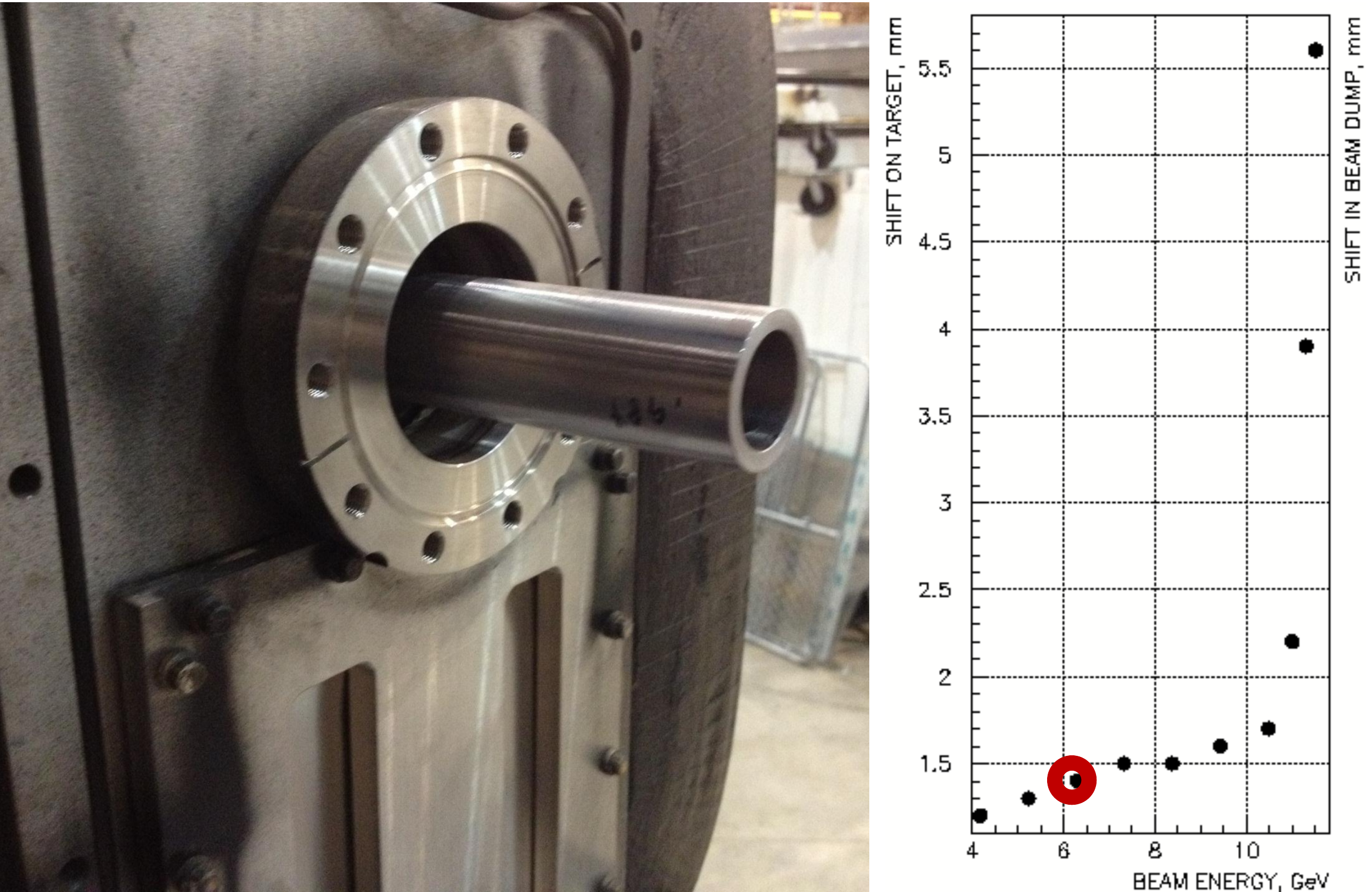}
   \label{fig:tosca}
\end{figure}

\begin{figure}[hbt]
   \caption [Moller: Beam shift after the dipole.]{Vertical beam
     shift on the BPM 1H04D downstream of the Hall A M\o{}ller
     polarimeter after the M\o{}ller dipole is turned ON.}
   \centering
        \includegraphics[width=1.0\textwidth]{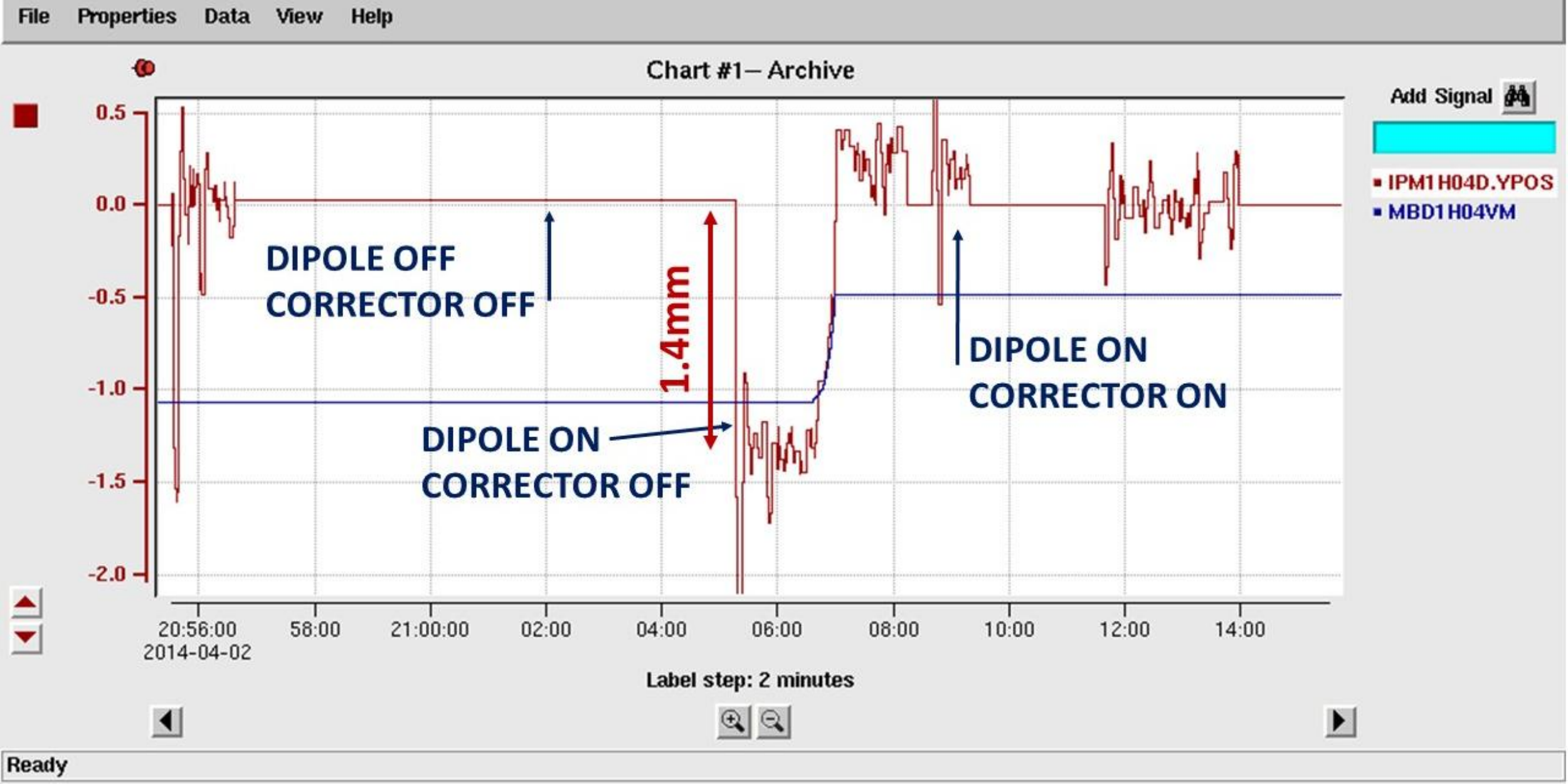}
   \label{fig:shift}
\end{figure}

\vspace{1em}
A critical part of the upgrade is a better
magnetic shielding of the electron beam line within the dipole magnet.
Operations at a higher beam energy require a higher field in the dipole,
saturating the existing shielding insertion. A steel pipe was added
into the shielding insertion throughout the length of the dipole, extending
from its upstream and downstream sides (see Fig.~\ref{fig:tosca}). 
Results of TOSCA calculations of the shielding properties of the M\o{}ller dipole
magnet after the upgrade is shown in Fig.~\ref{fig:tosca}. 
The residual magnetic field in the dipole is expected to cause
a 1.4~m downward deflection of a 6.05~GeV electron beam at the target location.
Fig.~\ref{fig:shift} shows the electron beam position shift
caused by the field in the M\o{}ller dipole magnet, in the area close
to the target. The shift of about 1.4~mm agrees well with 
the results of the TOSCA calculations.

The Hall A M\o{}ller polarimeter has two data acquisition and processing 
systems ~\cite{moldaq}:
\begin{itemize}
   \item Old DAQ is based on CAMAC, VME, NIM modules;
   \item New DAQ is based on VME module Flash-ADC F-250 designed by Jefferson Lab.
\end{itemize}

\begin{figure}[hbt]
   \caption [Moller: M\o{}ller DAQ's comparison.] {Result of the beam
     asymmetry measurement with two M\o{}ller polarimeter DAQs.}
   \centering
        \includegraphics[width=1.0\textwidth]{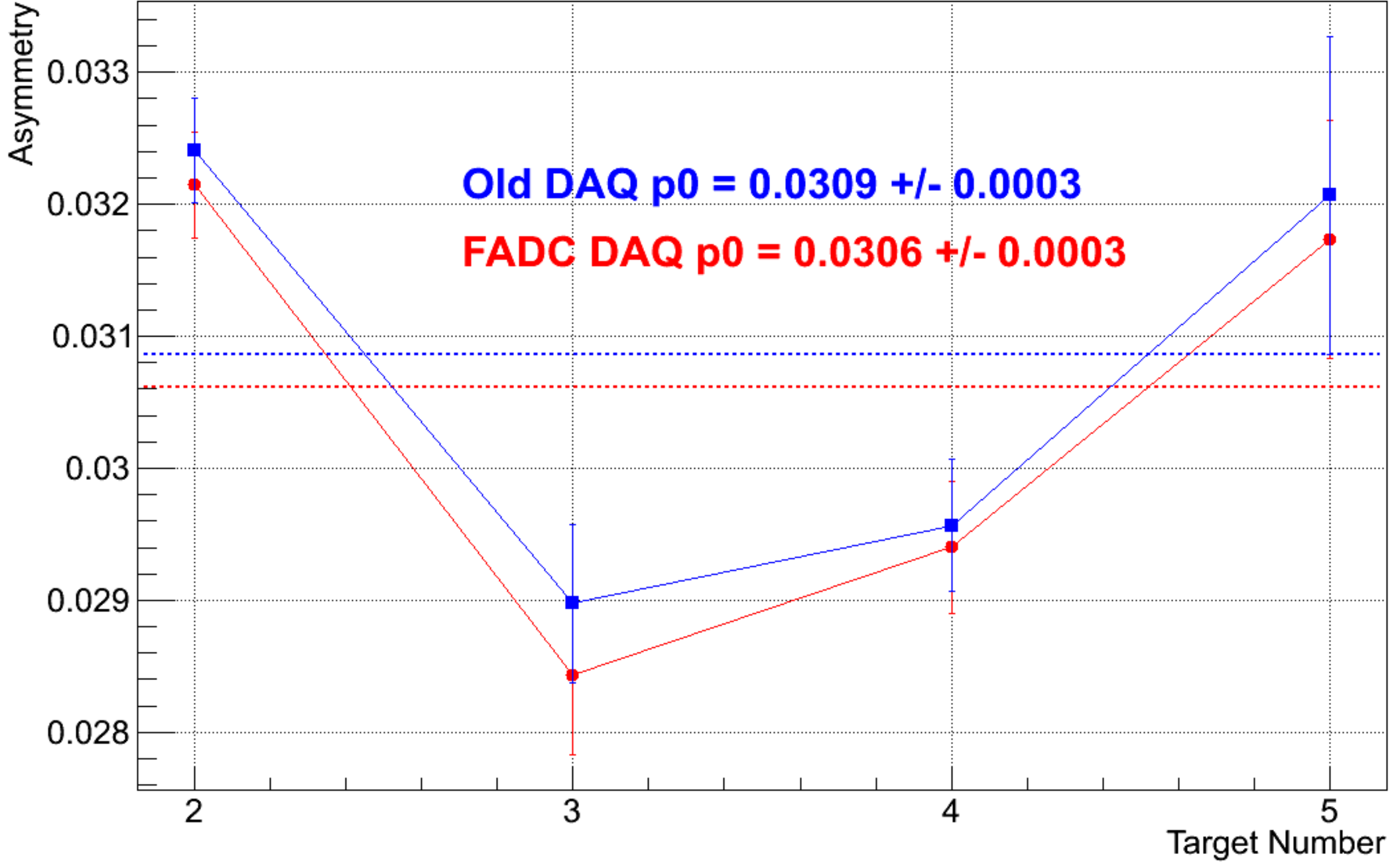}
   \label{fig:daqs}
\end{figure}

Both DAQs are used simultaneously to measure the electron beam
polarization. The old DAQ, in operation since 1998, is fully
functional but may not be repairable in case of malfunction, as the
electronic modules used are not in stock and are not manufactured anymore. The
new DAQ based on Flash-ADC, is in operation since 2010, is more
precise and provides more detailed data analysis. However, it
currently requires more careful adjustment and further
improvements. Fig.~\ref{fig:daqs} shows comparison of the asymmetry
values measured by both DAQs. Red dots show the measurement result
with the new DAQ system based on Flash-ADC, and blue dots show the
measurement result with the old DAQ system. The discrepancy between
the two DAQs results do not exceed the statistical error%
\footnote{Both DAQ systems collect largely overlapping event samples. 
The statistical errors should be strongly correlated.}%

\begin{figure}[hbt]
   \caption [Moller: Beam energy and position strip-chart.]
            {Strip-chart of the beam position stability (on the top)
              and the beam energy stability (on the bottom) at the
              time of the beam polarization measurement with the Hall
              A M\o{}ller polarimeter on December 14, 2014. }
   \centering
        \includegraphics[width=1.0\textwidth]{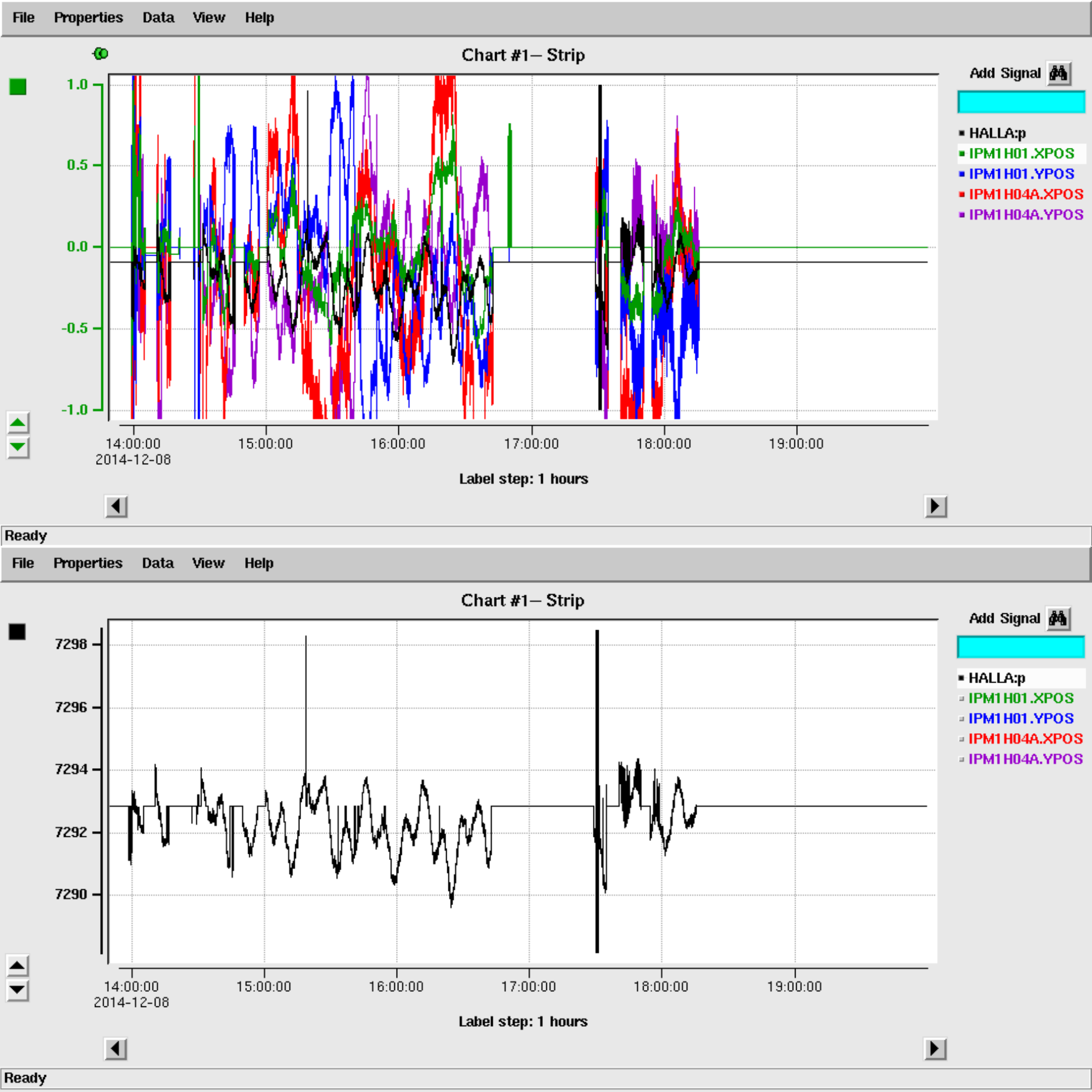}
   \label{fig:stripchart} 
\end{figure}

Another beam polarization measurement was done on December 14, 2014. The 
beam energy was about 7.4~GeV. Unfortunately, the beam energy and
position stability was very poor, as it is shown in
Fig.~\ref{fig:stripchart}, in comparison with the typical
stability of $\pm0.1$~mm during the 6~GeV era.  Therefore, no additional 
progress with commissioning of the Hall A M\o{}ller polarimeter or studying
the systematic errors was possible.

\subsubsection{Conclusion}
\label{sec:conclusion}

\begin{itemize}
   \item The Hall A M\o{}ller polarimeter is back to operation after
     the upgrade. Two beam polarization measurements were done in
     2014;
   \item Properties of the new shielding insertion to the M\o{}ller
     dipole magnet look consistent with the TOSCA calculations;
   \item A misalignment of the M\o{}ller polarimeter elements was
     found and partially fixed;
   \item Both old DAQ and FADC DAQ are working properly;
   \item More systematic studies of new polarimeter optics have to be
     done.
\end{itemize}


\clearpage




\subsection{Elastic study of Target boiling effect}

\hspace{1.5in}
D. W. Higinbotham, Dien Nguyen

\subsubsection{Introduction}

\begin{figure}[ht!]
\centering
\includegraphics[width=12cm]{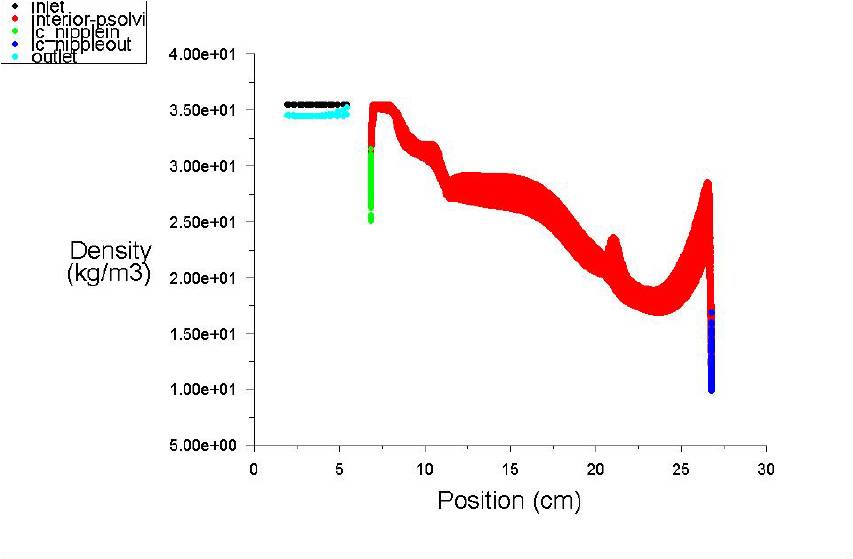}
\caption[Boiling: Simulation Density profile for gas $He4$ Target]{Simulation Density profile for gas $He4$ Target, pressure: $202 psia$, current: $120 \mu A$\label{overflow}}
\end{figure}

For cross section extractions, it is very important to know the absolute thickness of the targets.
In experiment E08014,  we used gas $He3$ and $He4$ targets.   From boiling effect studies of these targets,  we found large, current dependent target density fluctuations along the target cell.  
Silviu Covrig simulated this effect using his computational fluid dynamic model and was able to generate a density
profile similar to what was seen in the data (see Figure \ref{overflow}),\cite{Silviu}. 
In order to cross check that we are obtaining the correct absolute thickness, 
we checked to see if the $3He$ elastic peak was visible in the data for the lowest
Q2 in the data set.

\subsubsection{Elastic scattering}

Using a well measured reaction channel is often used as a cross check of absolute normalizations.    

For 3He, we have both good theoretical calculations and experimental measurements of the form factors and thus good knowledge of the cross sections \cite{J.S, I.S}.

\begin{figure}[ht!]
\centering
\includegraphics[width=10cm]{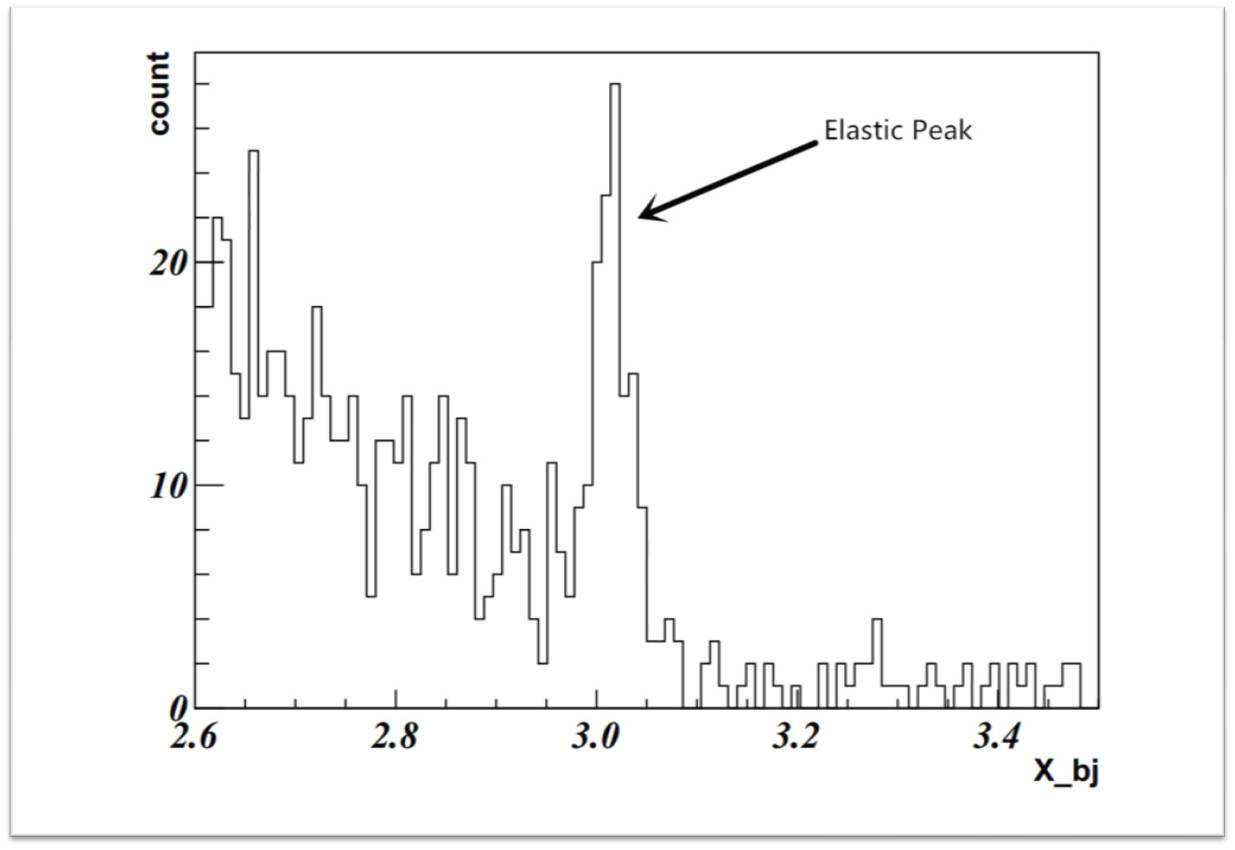}
\caption[Boiling: Elastic peak check from run 4074, experiment E08-014]{Elastic peak check from run 4074, experiment E08-014, $E0=3.356$ GeV, $\theta=21$ degrees
\label{f}}
\end{figure}

Using these results as a references, we have analyzed the elastic study of Target boiling effect for elastic scattering events (see Figure \ref{f} and \cite{Silviu}).

The elastic data is from runs with beam energy of 3.356 GeV, central scattering angle of 21 degrees, and central momemtum of 3.055 GeV.
Even with this high energy and large angle, we were still able to clearly identify the elastic peak using cuts on Cherenkov,
target vertex, trigger type, solid angle, and reaction point.

\subsubsection{Simulation}

Using the MCEEP simulation, we simulated the $3He$ elastic scattering  and applied the same cuts as were used in the data analysis.

To check that the code was working correctly, we first simulated a point target and checked the result against a hand calculation
and got good agreement.
Then to mimic the density profile, we took the extended target and broke
it into 4 sections with different densities and calculated the yield for each section. 
This of course can be repeat for a large number of sections (i.e. a finer grid of densities).  
The total yield is simply the sum of these yields.  After that we applied the radiative correction to total 
yield and compared to the yield of real data.

From the real data, one run yield is about 185 events.  We have five runs we can use to see the elastic peak and thus we 
expect to be able to extract about 1000 elastic events and thus we should be able to check the E08-014 density to the
three percent level.

Thus far, the results seem to be in reasonable agreement; but we still need to work the efficiency corrections which thus far
are still preliminary.

\subsubsection{Outlook}

We should be able to use this same technique to get check the target densities for the upcoming tritium experiments.  
With very first approximation the result look very promising. 

From simulation, we find that a run with an angle of $12$ and  $15$ degrees, beam energy of 2.2 GeV and a beam current of $25 \mu A$, active length of the target of $20$cm
would only need $\approx$ 1 hour of beam to achieve less then one percent uncertainty (see table1).
Thus, we should be able to check the target thickness of the tritium target to the 1-2 percent level.

\begin{table}[ht]

\caption{Yields and Statistic Errors in Percent}
\centering 
\begin{tabular}{ccccccc}
\hline\hline
 Target & Angle1 & Angle2 & Yield1 & Yield2 & Uncertainty1 & Uncertainty2\\
\hline
$He3$ & 12 & 15&  3e6 & 1.7e5 & 0.05 & 0.16\\
$H3$  & 12 & 15&  4e5 & 1.9e4 & 0.24 & 0.72\\
\hline
\end{tabular}
\label{table:nonlin}
\end{table}


\clearpage




\subsection{ARC Energy Measurement}

\begin{center}
contributed by Tong Su and D. Higinbotham
\end{center}

The ARC energy method uses a dipole magnet mapping system which is connected to a 9th dipole in series with the 8 dipoles
in the ARC along with beam position information to determine the CEBAF electron beam energy. Since the energy upgrade 
in the Hall A, the  excitation current range for the dipoles has been extended and all of the dipole were refurbished.   
Thus for the 12GeV era, dipoles need to re-calibrated and the beam energy vs. ARC current set points needs to be redetermined. 

\begin{figure}[htbp]
\centering
\includegraphics[width=\linewidth]{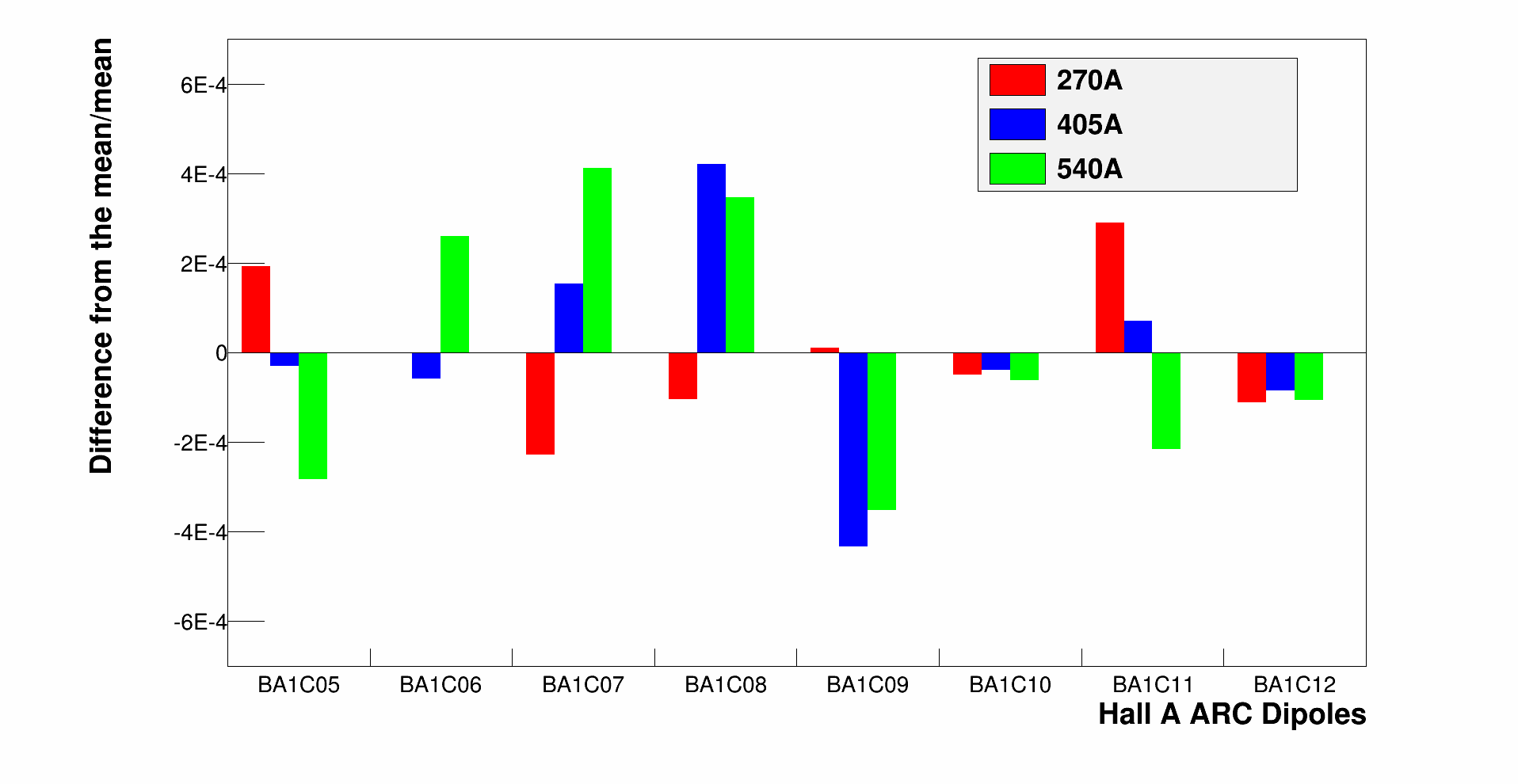}
\caption[ARC: The deviation from the mean of the 8 dipoles in the Hall A ARC]{The deviation from the mean of the 8 dipoles in the Hall A ARC.\label{BL-deviation}}
\end{figure}

The absolute field in the 9th dipole is calibrated using a NMR probe and the field integral is be determined with coils 
mount on a translation table that moves the coils through the magnetic field at a known velocity.  The field integral of
the nine dipoles has been measured w.r.t to the 8 dipoles down in the ARC.   The field integral difference of the 8 ARC 
dipoles at currents 270A, 405A, 540A is shown in Fig.~\ref{BL-deviation}.

\begin{figure}[htbp]
\centering
\includegraphics[width=\linewidth]{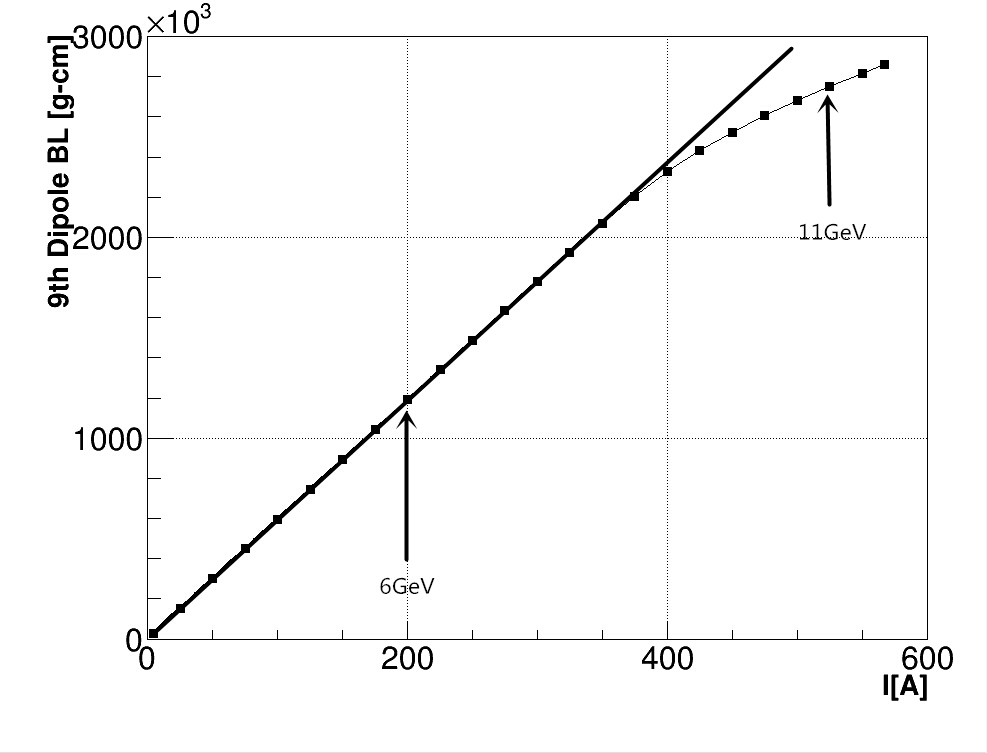}
\caption[ARC: The relations between the field integral and current for the 9th magnet mapper dipole.]{The relations between the field integral and current for the 9th magnet mapper dipole.\label{BL-vs-I}}
\end{figure}

For the 9th dipole, field integral at the full excitation current range( up to 570A) has been measured and as one can
see in Fig.~\ref{BL-vs-I} at the highest excitation currents the magnet is starting to saturate as can be seen from
the  field vs. current has deviating from linearity.

To make sure that the set current of accelerator is equal to the true current in the dipoles, we have installed
a calibrated Ultrastab Saturn unit with a 2000A head.   With this unit in place, we found an issue with the 
calibration constant that was being used by the IOC reporting the ARC current.  Once this was fixed in late 
December, the current set by accelerator and our independent Ultrastab measurements have agreed at better then
the 1E-4 level.  

Using the 9th dipole and its relation to the other 8 dipoles allows one to determine the field integral of the dipoles in the ARC.
Combining this information with the bending angle from the Hall A scanners, one can determine the absolute energy to the $5x10^{-4}$ level.  By changing ARC tune to a fully dispersive mode, instead of its normal tune and using the super-harps to determine the
deflection angle this precision can be pushed to the few $10^{-4}$ level.  In the future, we will compare the results of the ARC energy measurements with the energy determined using spin precession~\cite{Grames:2004mk,Higinbotham:2009ze}.

We do note that in order to get the precise energy for the beam on target, for both these measurement techniques the synchrotron 
needs to be taken into account.   To do this precisely, we have been working with Yves Roblin who within the context of his 
model of the accelerator can precisely calculate that correct for the exact geometry of the accelerator.



\clearpage

\section{Future Experiments}

\subsection{E14009: Ratio of the electric form factor in the mirror nuclei $^3$He and $^3$H}\label{sec:e14009}

\begin{center}
contributed by L. S. Myers, D. W. Higinbotham and J. R. Arrington.
\end{center}

The E14009 experiment~\cite{ref0} proposes to measure the ratio of the 
electric form factors of $^3$He and $^3$H over a range of $Q^2$
from 0.05 to 0.09~GeV$^2$. From this measurement, the relative
charge radii for the A=3 nuclei can be determined. From this 
relationship, the experiment plans to extract the charge radius
difference with an uncertainty of $\sim$15\% -- a significant
reduction from the current 50\% uncertainty. The JLab PAC42 
approved the experiment for the requested 1.5 days of beam time. 
This experiment will run with at the same time as the other 
approved triton experiments~\cite{ref1,ref2,ref3}. 

\subsubsection{Experimental Layout}\label{sec:experimental_layout}

\begin{figure}[hbt]
  \caption[E14009: Setup for E14009.]{A schematic diagram of the
    experimental setup for E14009. Only the left HRS will be 
    employed for this measurement. Targets of $^{1,2,3}$H, $^3$He,
    and $^{12}$C are planned for study. A custom collimator 
    plate (see Fig.~\ref{fig:plate}) will be utilized.}
  \centering
  \includegraphics[width=\textwidth]{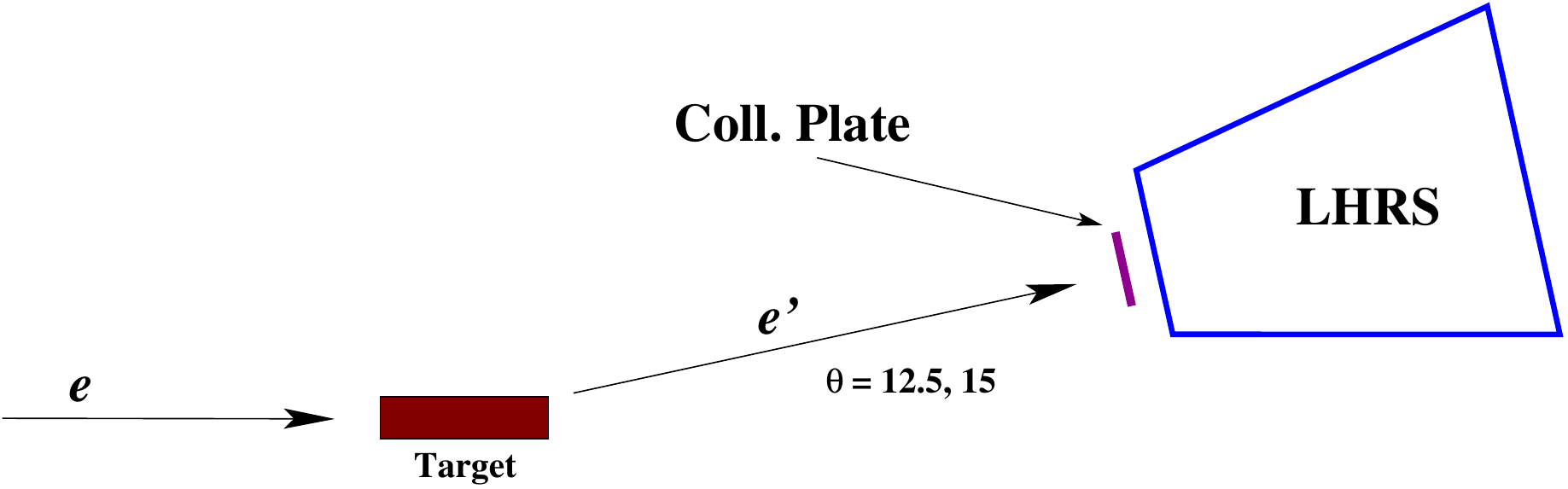}
  \label{fig:setup}
\end{figure}

The experiment will be a single-arm measurement of elastic electron
scattering from $^3$He and $^3$H (see Fig.~\ref{fig:setup}). It 
will utilize the left high-resolution spectrometer (LHRS) in 
Hall A positioned at 12.5$^\circ$ and 15.0$^\circ$. In general, the 
setup is very similar to previous form factor measurements in Hall A,  
as well as the other planned triton experiments.

There are two unique features of the setup for E14009. First, the 
electron beam will be set to 1.1~GeV (the lowest achievable energy 
with the upgraded accelerator) and the current will be limited to 
only 5~$\mu$A. The energy is necessary to get down to $Q^2$ of 
0.05--0.09~GeV$^2$. At these momentum transfers, the cross sections 
and scattering rates are large. The reduced current eliminates the 
need for a large trigger prescale factor in the data acquisition.

\begin{figure}[hbt]
  \caption[E14009: Custom collimator plate.]{The custom
    collimator plate for the experiment (not to scale). The 
    bin labeled '1' is the smallest $Q^2$ point at the given 
    angle setting. The holes above and below the center slot 
    are for simultaneous optics measurements.}
  \centering
  \includegraphics[width=0.5\textwidth]{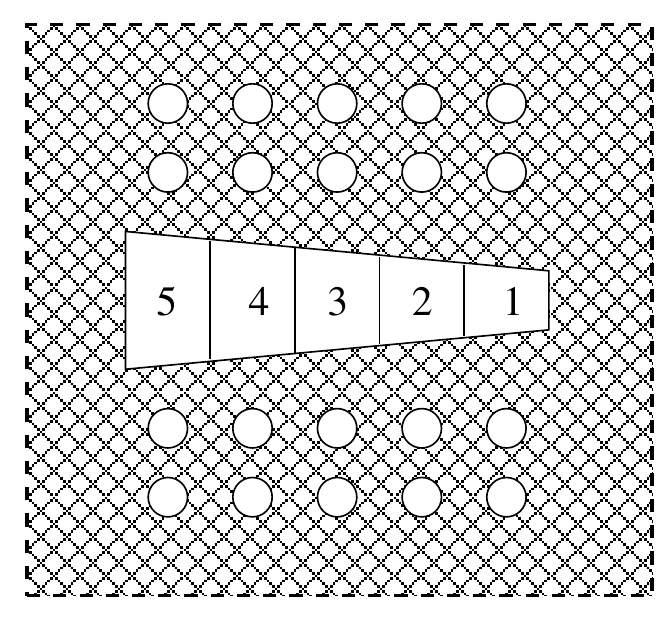}
  \label{fig:plate}
\end{figure}

The other unique feature of E14009 is the custom collimator plate 
(Fig.~\ref{fig:plate}) to be placed on the spectrometer. This plate 
serves three purposes: (1) the center slot further reduces the event 
rate in the HRS while (2) the tapered design reduces the rate on the 
low-angle side thereby balancing the rates in each of the five $Q^2$ 
bins, and (3) the rows of holes above and below the center slot allow 
for in situ optics measurements.

\subsubsection{Anticipated Results}\label{sec:anticipated_results}

The count rates for the elastic electrons are expected to be
$\sim$10$^5$ counts/hr per $Q^2$ bin. Data taking for each target will
take 1.5~hrs at each HRS angle. As a result, the systematic
uncertainties ($\sim2\%$) will dominate the statistical ($\textless$0.5\%). 
Within the systematic uncertainty, the largest contribution is 
expected to come from the relative thickness of the $^3$H and $^3$He 
targets. This uncertainty ($\sim$1.5-2.0\%) will be finalized using
deep inelastic scattering from both targets during the running of 
the MARATHON experiment.

\begin{figure}[hbt]
  \caption[E14009: Anticipated results.]{The expected results
    of the measurement of the charge form factor ratio of $^3$He 
    and $^3$H. The $^3$He radius is fixed to 1.96~fm~\cite{ref4}. The solid 
    line assumes a $^3$H radius of 1.76~fm with $\pm$0.03~fm
    used to calculate the dashed lines.}
  \centering
  \includegraphics[width=0.75\textwidth]{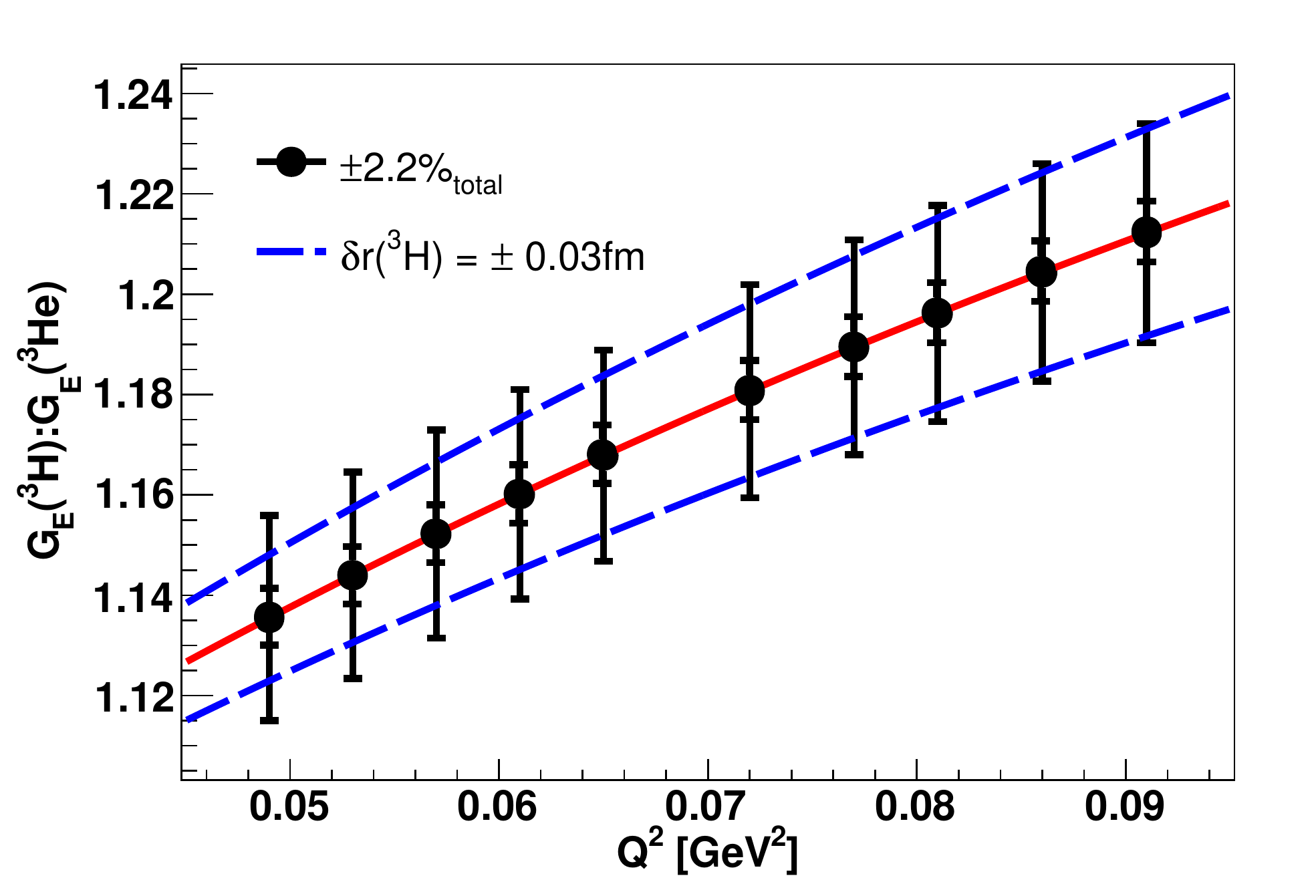}
  \label{fig:results}
\end{figure}

The anticipated results for the experiment are shown in Fig.~\ref{fig:results}.
Assuming a conservative estimate of the uncertainties, the $^3$He--$^3$H
can be determined to $\sim$0.03~fm, a reduction of 70\% from the 
current uncertainty. The model dependence of the radius extraction
is constrained by examining the evolution of the form factor ratio
over the full range of $Q^2$. 

This measurement will also be able to extract the most precise
value of the $^3$H charge radius by using $^3$He radii available
from isotopic shift measurements~\cite{ref5,ref6} and soon from 
muonic $^3$He~\cite{ref7}. These combined results will allow for 
the extraction of the proton and neutron radii in the A=3 nuclei
which can be compared to ab initio calculations.


\clearpage

\subsection[Super Bigbite Spectrometer (SBS)]{Super Bigbite Spectrometer}
\label{sec:sbs}

\begin{center}
    Progress by the Super Bigbite Collaboration
\end{center}

\begin{center}
contributed by S.~Riordan \\
for the Super Bigbite Collaboration.
\end{center}

\subsubsection{Overview}\label{sec:sbs:overview}

The Super Bigbite (SBS) project is a collection of experiments based upon utilizing large-acceptance single dipole spectrometers designed to operate in a high rate environment in conjunction with the upgraded $12~\mathrm{GeV}$ CEBAF accelerator.  The official DOE project of SBS focuses around three coincident nucleon elastic form factor measurements at high $Q^2$, which require large acceptance but moderate momentum resolution.  In addition, there are two approved experiments which also utilize this equipment, making the SBS collaborative effort contain

\begin{itemize}
    \item{E12-07-109, GEp}
    \item{E12-09-016, GEn}
    \item{E12-09-019, GMn}
    \item{E12-09-018, SIDIS}
    \item{E12-06-122, $A_1^n$}.
\end{itemize}

The collaboration underwent a successful annual review of the project in November, meeting the recommendations of the previous review.  The full report can be found online~\cite{sbs:review}.  More documentation over the entirety of the project can be found online~\cite{sbs:webpage}.

\subsubsection{Instrumentation Progress}

\paragraph{48D48 Magnet and Beamline}

One critical component of the SBS project is the 48D48 magnet, which serves as the magnetic element for the hadronic arm of these experiments.  In the last year, progress was made in completing drawings for the assembly in the SBS configuration, the modification of the iron yoke (completed in May), and the construction of the new racetrack and saddle coils and support platform.  The magnet floor layout was changed from beam-left to beam-right to ease installation and compatibility between experiments.

The new racetrack coils have been fabricated and delivered by vendors to Jefferson Lab and underwent local testing.  The saddle coil is under construction and is expected to be delivered in July 2015.  The power supply was delivered in October and water cooling components were acquired and assembled.   A test of the magnet at 200~A was performed and compared to TOSCA simulations with a 2\% deviation observed with further studies underway.  

In addition, beamline corrector magnets and shielding configurations are under study to compensate for fields as the exit beamline passes through the magnet iron.  The scattering chamber vacuum snout is anticipated to arrive in February with the new consideration of magnet beam-right.  Simulations of the background for the shielding has been underway with support of the Yerevan group.

\paragraph{GEM Detectors and Tracking Hardware}

Several sets of GEM detectors are being constructed by groups at INFN and UVA and will be used in both the hadron and electron arms of these experiments.  In 2014 major milestones were completed in the GEM production.  The UVA group has received the parts for GEM construction and has now assembled seven modules which are undergoing testing.  A high intensity X-ray source has been constructed to provide the capability to do high background testing with rates comparable to SBS conditions.

The INFN collaboration has assembled eight $40\times50~\mathrm{cm}^2$  modules, four of which are undergoing testing with cosmics and HV curing.  Short, opportunistic in-beam tests at COSY were performed in December to study the response with a high-intensity proton beam and study HV and gas flow.

A review over the choice of electronics, either the SRS or INFN VME-based MPD APV25 readout systems, was carried out. The MPD system was selected for future use.

The INFN group is also working on the development of two small silicon microstrip planes ($10\times20~\mathrm{cm}^2$, $50~\mathrm{\mu m}$ pitch) to improve the tracking of the primary particles in the front tracker.  Evaluation of the response of this system and electronics using a prototype is ongoing and assembly and testing of the detector is expected in 2015.

\paragraph{ECal}


A primary concern for the electromagnetic calorimeter is the darkening of lead-glass blocks as they absorb radiation and a thermal annealing method which operates continuously has been under investigation at JLab. In 2014 tests were performed quantifying the restoration capabilities using blocks that had been irradiated by Idaho State. The results showed that such a method was feasible but also that when held at a high temperature the optical transparency was slightly reduced.  Simulations were performed which quantified the light yield for electromagnetic showers with a transparency gradient to ensure that sufficient resolution would be maintained.  A 16-block prototype has been assembled with support from the Yerevan group at Jefferson Lab waiting to undergo testing in the hall when beam is restored in spring using a coincident trigger to select an electron with well defined energy.  Stony Brook University has agreed to lead the development of a 200-block prototype with drawings produced with the support of the Yerevan group.  This prototype will have the design finalized and constructed in 2015 for the purposes of testing a mechanical and thermal design and costing.

\paragraph{HCal-J}

The hadronic calorimeter, HCal-J, serves as a primary component in hadron detection for the form factor and SIDIS experiments and is being constructed at Carnegie Mellon University with contributions of funding and manpower by INFN/Catania.  In 2014 a fully working prototype was constructed and tested with cosmics.  These tests were compared with detailed Monte Carlo simulations, in particular to ensure sufficient timing resolution required for the $G_E^n$ measurement.  Iron, scintillator, and wavelength shifter for the full detector has been received and are undergoing machining and assembly.  Design of the light guides to optimize readout is undergoing testing using a three-piece acrylic design and transmission tests with comparisons to simulation are ongoing.

\paragraph{Polarized ${}^3\mathrm{He}$ Target}

The polarized ${}^3\mathrm{He}$ target is at the heart of the $A_1^n$, $G_E^n$, and SIDIS experiments and provides an effective polarized neutron target.   For SBS beam currents,  a design that includes metal end-cap windows, a convection design, and two laser pumping chambers is being developed at UVA.  A target-cell design for for this was selected in October and after fabrication, simulated-beam tests are planned.

\paragraph{Gas Cherenkov}

A 600 PMT gas Cherenkov detector to be used for electron identification is being developed jointly by collaborators at William and Mary, North Carolina A\&T State University, University of Glasgow, and James Madison University.  The mirror assembly has been produced by the William and Mary machine shop and the optics configuration has been tested prompting the order of the final mirrors.  The pressure vessel design is underway at Jefferson lab.  NC A\&T are in the procurement process for the PMT array and magnetic shielding.  JMU has characterized 600 of 800 PMTs and is grouping them based on gains.

\paragraph{Coordinate Detector}

A scintillator coordinate detector will serve as a hodoscope to determine the electron position in front of the electromagnetic calorimeter in the $G_E^p$ experiment.  The position resolution of such a detector will allow for a much cleaner identification of proton elastic events than by the calorimeter alone by exploiting the specific electron-hadron kinematic correlation.  Efforts led by the Idaho State and Saint Mary's group for construction are underway after the completion and testing of a prototype module.  In 2014, a sample set of scintillator bars were ordered and procured from the Fermilab extrusion facility and considerations are underway to determine where machining will be done.  Full production of the scintillator is expected at the end of February.  Sample waveshifting fibers from St. Gobain have been received at Jefferson Lab for testing.  Drawings for the support structure have been completed and vendors are being selected for fabrication.  Electronics are also undergoing testing using a NINO chip and FASTBUS design for readout.

\paragraph{RICH}

A RICH detector is included with the SIDIS experiment for PID and work of refurbishment of an existing detector is being led by the University of Connecticut group.  The detector has been delivered there from storage at UVA.  Detailed representations have been included in the Monte Carlo simulation and an analysis has been performed in the reconstruction and PID capabilities by the same group.

\paragraph{Timing Hodoscope}

The Bigbite timing hodoscope consists of 90 600x25x25mm plastic scintillator bars, each equipped with 2 fast PMTs and is necessary for accurate coincident time of flight measurements.  Tests performed at Glasgow show that 100~ps timing resolution is possible.  The bars and lightguides for its construction have been procured and the lightguides have been glued to the scintillator.  Glasgow is also developing a NINO-based frontend amplifier/discriminator for this component, as well as for the GRINCH and coordinate detector.

\paragraph{Software and Simulation}

The Geant4 and ROOT-based SBS Monte Carlo simulation, {\verb g4sbs }, has undergone a huge amount of development with contributions from many different groups.  A full realization of all the detector systems has been incorporated, including a detailed representation of the magnets including field clamps, beamline including shielding and magnetic elements, target chamber, RICH, GRINCH, and ECal.  Optical photon processes are included to evaluate light yields and backgrounds in the detector responses.  A full evaluation of the trigger rates including background for the three form factor measurements has been performed including the calorimeter responses for various particle types and coincident trigger logic.  Plans are underway to incorporate the simulation into the analysis framework to test various reconstruction algorithms on pseudodata.   The INFN group has been developing a new neural network-based tracking algorithm in parallel of the existing tree search method and will be incorporated.

\clearpage

\section{Summaries of Experimental Activities}

\subsection[E05102: Double polarized asymmetries
in quasi-elastic processes ${}^3\vec{\mathrm{He}}\left(
\vec{\mathrm{e}},\mathrm{e'} \mathrm{d }\right)$ and 
${}^3\vec{\mathrm{He}}\left(\vec{\mathrm{e}},\mathrm{e'} 
\mathrm{p}\right)$]{E05102: Measurement of double polarized asymmetries
in quasi-elastic ${}^3\vec{\mathrm{He}}\left(
\vec{\mathrm{e}},\mathrm{e'} \mathrm{d }\right)$ and 
${}^3\vec{\mathrm{He}}\left(\vec{\mathrm{e}},\mathrm{e'} 
\mathrm{p}\right)$}\label{sec:e05102}

\begin{center}
Contributed by M.~Mihovilovi\v{c} and S.~\v{S}irca.
\end{center}

\subsubsection{Introduction}\label{sec:intro}
The E05-102 experiment~\cite{e05102} is dedicated to the study of the ${}^3\mathrm{He}$ nucleus  
by measuring double-polarization (beam-target) transverse and longitudinal 
asymmetries in quasi elastic processes 
${}^3\vec{\mathrm{He}}\left(\vec{\mathrm{e}},\mathrm{e'} \mathrm{d }\right)$ and  
${}^3\vec{\mathrm{He}}\left(\vec{\mathrm{e}},\mathrm{e'} \mathrm{p}\right)$ as 
a function of the missing momentum $(p_{\mathrm{miss}})$. The purpose of this 
measurement is to challenge the existing predictions on the structure and properties 
of this three-nucleon system and to provide
new constraints for the theoretical models. In particular, the 
${}^3\vec{\mathrm{He}}\left(\vec{\mathrm{e}},\mathrm{e'} \mathrm{d }\right)$ reaction
is a uniquely sensitive probe of hadron dynamics in ${}^3\mathrm{He}$ and the 
isospin structure of the underlying electro-magnetic currents, while the 
proton channels ${}^3\vec{\mathrm{He}}\left(\vec{\mathrm{e}},\mathrm{e'} \mathrm{p}\right)\mathrm{d}$
and ${}^3\vec{\mathrm{He}}\left(\vec{\mathrm{e}},\mathrm{e'} \mathrm{p}\right)\mathrm{pn}$
give us the ability to precisely study final-state-interactions and 
nucleon-nucleon correlations.  The experiment was performed in Summer 2009 
by using the polarized ${}^3\mathrm{He}$ target and a
High Resolution Spectrometer (HRS-L) in coincidence with the large-acceptance 
spectrometer BigBite~\cite{miha_NIM}.

\subsubsection{The deuteron channel}
Because of a cleaner experimental signature and more profound implications on the structure
of ${}^3\mathrm{He}$ our efforts have primarily been focused on the analysis of the 
${}^3\vec{\mathrm{He}}\left(\vec{\mathrm{e}},\mathrm{e'} \mathrm{d }\right)$ channel, 
which is now complete and published~\cite{eed_PRL}. The measuremens have 
been confronted with the state-of-the-art Faddeev calculations and the comparison 
of the asymmetries shows a fair agreement in terms of their functional dependencies in
$p_{\mathrm{miss}}$ and $\omega$, but significant discrepancies remain, in particular 
beyond the quasi-elastic peak.

\begin{figure}[!ht]
\begin{center}
\begin{minipage}[t]{0.6\textwidth}
\hrule height 0pt
\includegraphics[width=\textwidth]{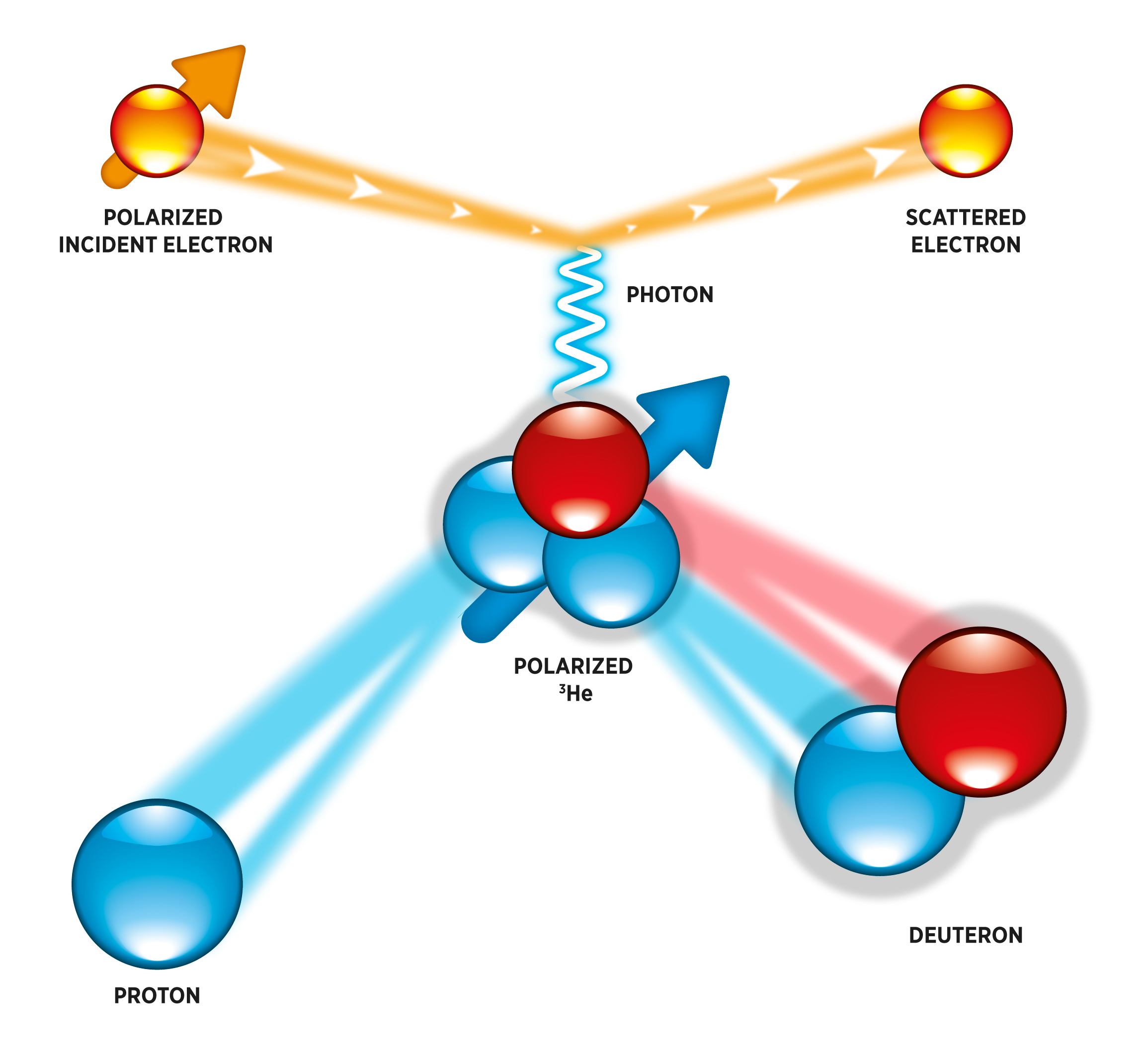}
\end{minipage}
\hfill
\begin{minipage}[t]{0.35\textwidth}
\hrule height 0pt
\vskip -3mm
\caption[E05102: Schematic of ${}^3\vec{\mathrm{He}}\left(\vec{\mathrm{e}},\mathrm{e'} \mathrm{d }\right)$]{Schematic drawing of the ${}^3\vec{\mathrm{He}}\left(\vec{\mathrm{e}},\mathrm{e'} \mathrm{d }\right)$
reaction. Longitudinally polarized electrons are scattered off a polarized ${}^3\mathrm{He}$ target. In the process the nucleus
breaks up into a proton and a deuteron, which is detected in coincidence with the scattered electron.
\label{fig:e05102:eed}}
\end{minipage}
\end{center}
\end{figure}

\begin{figure}[!ht]
\begin{center}
\begin{minipage}[t]{0.68\textwidth}
\hrule height 0pt
\includegraphics[width=\textwidth]{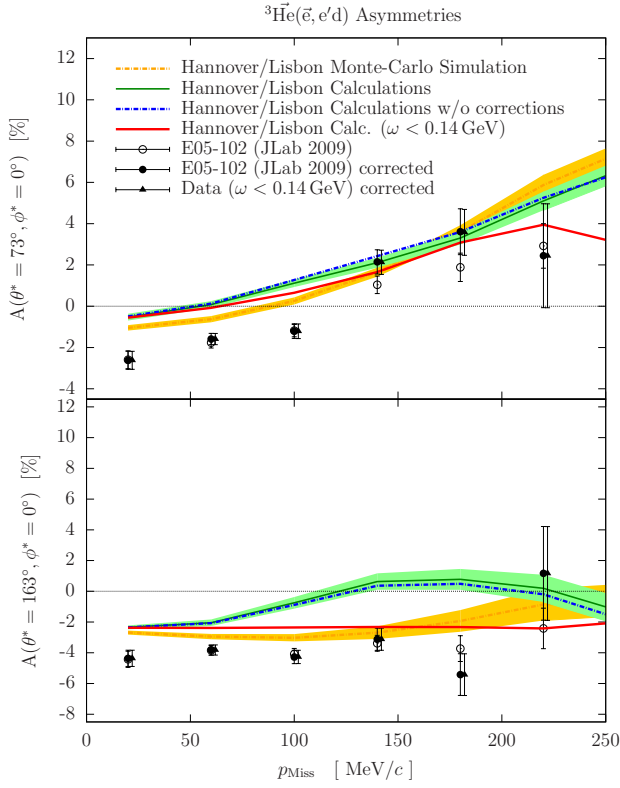}
\end{minipage}
\hfill
\begin{minipage}[t]{0.3\textwidth}
\hrule height 0pt
\vskip -3mm
\caption[E05102: Asymmetries in quasi-elastic ${}^3\vec{\mathrm{He}}\left(\vec{\mathrm{e}},\mathrm{e'} \mathrm{d }\right)$]{The longitudinal (top) and transverse (bottom) asymmetries in the quasi-elastic
process ${}^3\vec{\mathrm{He}}\left(\vec{\mathrm{e}},\mathrm{e'} \mathrm{d }\right)$ as a 
function of missing momentum $(p_{\mathrm{miss}})$. The empty circles represent data points
measured in the E05-102 experiment. Full circles show the same results corrected for
the effects of bin migration, which arise due to radiative losses. The full triangles
show the measured asymmetries for the events from the vicinity of the quasi-elastic peak
$(\omega<0.14\,\mathrm{GeV})$. The  full green line and the blue dash-dotted line show
the acceptance averaged theoretical calculations by the Hannover-Lisbon group with and 
without radiative corrections, respectively. The red line shows the corrected theoretical prediction
for the quasi-elastic peak. The yellow dashed line represents the results of a stand-alone 
Monte-Carlo simulation. 
\label{fig:e05102:asymmetries}}
\end{minipage}
\end{center}
\end{figure}

To ensure a truthful comparison of theory to the data, shown in Fig.~\ref{fig:e05102:asymmetries}, 
a sophisticated algorithm for interpolating and averaging a discrete mesh of theoretical 
points over the measured kinematic acceptance was developed~\cite{HallA_2013}. Additionally, 
a stand-alone Monte-Carlo simulation was designed to simulate the physics process under scrutiny. 
The program considers the cross-sections provided by the theoretical groups, radiative corrections 
in terms of the peaking approximation and the spectrometer acceptances, defined by their collimator 
sizes. This simulation gave us the ability to validate the theoretical asymmetries 
determined by the averaging algorithm and to estimate and correct the 
effects of bin migration, which arise due to radiative losses.

\subsubsection{The proton channels}
\begin{figure}[!ht]
\begin{center}
\begin{minipage}[t]{0.68\textwidth}
\hrule height 0pt
\includegraphics[width=\textwidth]{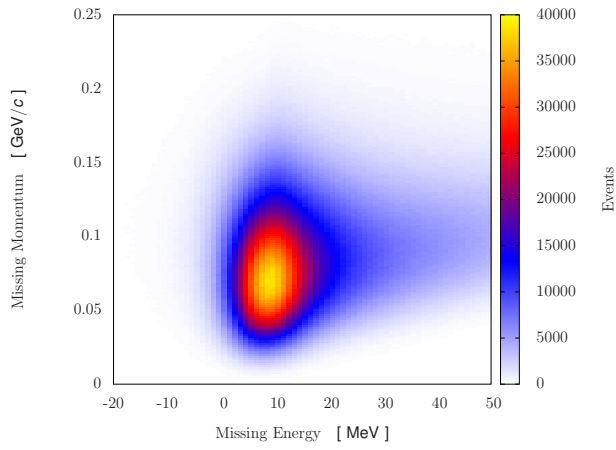}
\end{minipage}
\begin{minipage}[t]{0.3\textwidth}
\hrule height 0pt
\vskip -3mm
\caption[E05102: Distribution of measured ${}^3\vec{\mathrm{He}}\left(\vec{\mathrm{e}},\mathrm{e'} \mathrm{p }\right)$ 
events]{The distribution of measured ${}^3\vec{\mathrm{He}}\left(\vec{\mathrm{e}},\mathrm{e'} \mathrm{p }\right)$ 
events as a function of missing energy $(E_{\mathrm{miss}})$ and missing momentum $(p_{\mathrm{miss}})$. Due to insufficient
resolution the two-body breakup peak at $E_{\mathrm{miss}}=5.5\,\mathrm{MeV}$ can not be separated
from the three-body breakup at $E_{\mathrm{miss}}=7.7\,\mathrm{MeV}$. Additionally, the distribution
has a long tail in $E_{\mathrm{miss}}$ which needs to be interpreted properly for a objective 
comparison of the data to the theoretical predictions. 
\label{fig:e05102:proton}}
\end{minipage}
\end{center}
\end{figure}

The analysis of the two proton channels is ongoing. Due to insufficient resolution of
the experimental apparatus, the two- and three-body channels in the proton knockout 
processes can not be disentangled~\cite{mihaPhD}. Hence, the measured asymmetries are the sums of
two reaction processes, which makes the comparison of the data to the simulation 
much more challenging. To overcome this obstacle, the calculated asymmetries for 
the two processes at each kinematic point will be weighted by the predicted 
cross-section and then summed. For the interpolation and averaging of the 
available theoretical calculations, a similar approach as for the deuteron channel 
will be employed, but with an additional degree of freedom for the three-body proton 
breakup, where the $p-n$ system is no longer bound, thus allowing for the final states with 
different missing energies $E_{\mathrm{miss}} \geq 7.7\,\mathrm{MeV}$ and consequently 
different asymmetries (see Fig.~\ref{fig:e05102:proton}). It is foreseen that this last 
part of the analysis will be done in 2015 and the results sent to the publication by mid-2016.

\clearpage

\subsection{E07006: Short Range Correlations (SRC)}\label{sec:e07006}

\begin{center}
contributed by N.~Muangma and V.~Sulkosky
\end{center}


Nucleon-Nucleon Short Range Correlations (NN-SRC) have been studied in both 
the inclusive reaction $(\rm{e},\rm{e'})$ and the exclusive triple-coincidence 
reaction.  For experiment E07-006, we are analyzing two reaction channels:  
the exclusive reaction $(\rm{e},\rm{e'pN})$ and the semi -inclusive reaction 
$(\rm{e},\rm{e'p_{recoil}})$.  This experiment is a continuation of the first Hall A 
SRC experiment, E01-015~\cite{Shneor:2007tu,Subedi:2008zz,Monaghan:2013ah} by extending
the measurement to higher missing momenta.  The experimental details and the final results 
of the exclusive reaction can be found in~\cite{Korover:2014dma}. 

The semi-inclusive channel is an attempt to study NN-SRC, in which the statistics 
are significantly better than the exclusive triple-coincidence reaction.  We are
exploring if the backward tagged protons, after subtraction of the random background events, 
are coming from the SRC NN-pair, where the undetected nucleon carried the transferred 
momentum.   

The electrons were detected in the L-HRS, and the recoil protons were detected in the 
BigBite spectrometer, which was implemented with the hadron detector 
package~\cite{Mihovilovic:2012hi}.  The recoil protons were detected at much larger 
angle compared to the transferred momentum $\vec{q}$.  By using the pion rejectors in 
the L-HRS, the separation between the pions and electrons was very clean.  At this 
stage of the analysis, we are also extracting the inclusive ratios of the deuteron, $^{4}$He 
and $^{12}$C versus $x_{bj}$ to verify that they are consistent the measured $a_{2}$ values, where
$a_{2}$ is related to the number of 2N correlations in the nucleus relative to that of the 
deuteron. 

The backward protons detected in BigBite are identified with the measured momentum from 
the multi-wire drift chambers (MWDC) and the energy deposited in a pair of scintillator 
planes, known as the dE and E planes.  In principle, the protons can be identified directly
from the dE and E planes; however at high momenta, the protons do not deposit enough energy 
in the dE plane, which requires the use of the the MWDC and the E plane for proton PID.

\begin{figure}[!h]
\begin{center}
\begin{minipage}[t]{0.68\textwidth}
\hrule height 0pt
\includegraphics[width=\textwidth]{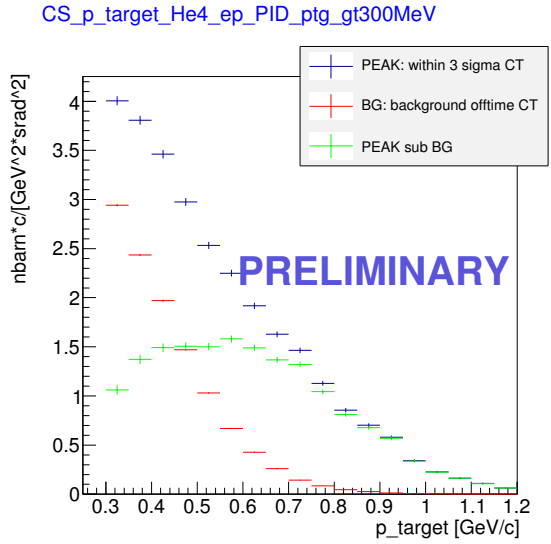}
\end{minipage}
\hfill
\begin{minipage}[t]{0.3\textwidth}
\hrule height 0pt
\vskip -3mm
 \caption[E07006: Recoil Proton Momentum Distribution.]{The six-fold differential 
    cross section for the $^{4}\rm{He}(\rm{e},\rm{e'p_{recoil}})$ reaction versus the detected 
    recoil proton momentum, $p_{\rm{target}}$.  The green data points represent the data 
    after subtraction of the random background.}
  \label{fig:e07006-ptg}
\end{minipage}
\end{center}
\end{figure}

To obtain the semi-inclusive data above the random background distribution, we are using 
two coincidence time cuts: the first with a cut on the coincidence time-of-flight (CTOF) 
peak, and the second with a cut to the sides of the CTOF peak.  For the momentum distribution, 
the random background is estimated from the off-peak CTOF cut.  Then the data above background 
is the data within the CTOF peak subtracted from the estimated random background.   
Figure~\ref{fig:e07006-ptg} shows the six-fold differential cross section versus the proton 
momentum, which has been corrected for energy losses.  The blue data represent the signal and 
background together, the red shows the estimated background contribution, and the green is the 
background subtracted distribution.

\begin{figure}[!h]
\begin{center}
\begin{minipage}[t]{0.68\textwidth}
\hrule height 0pt
\begin{tabular}{c}
  \includegraphics[width=\textwidth]{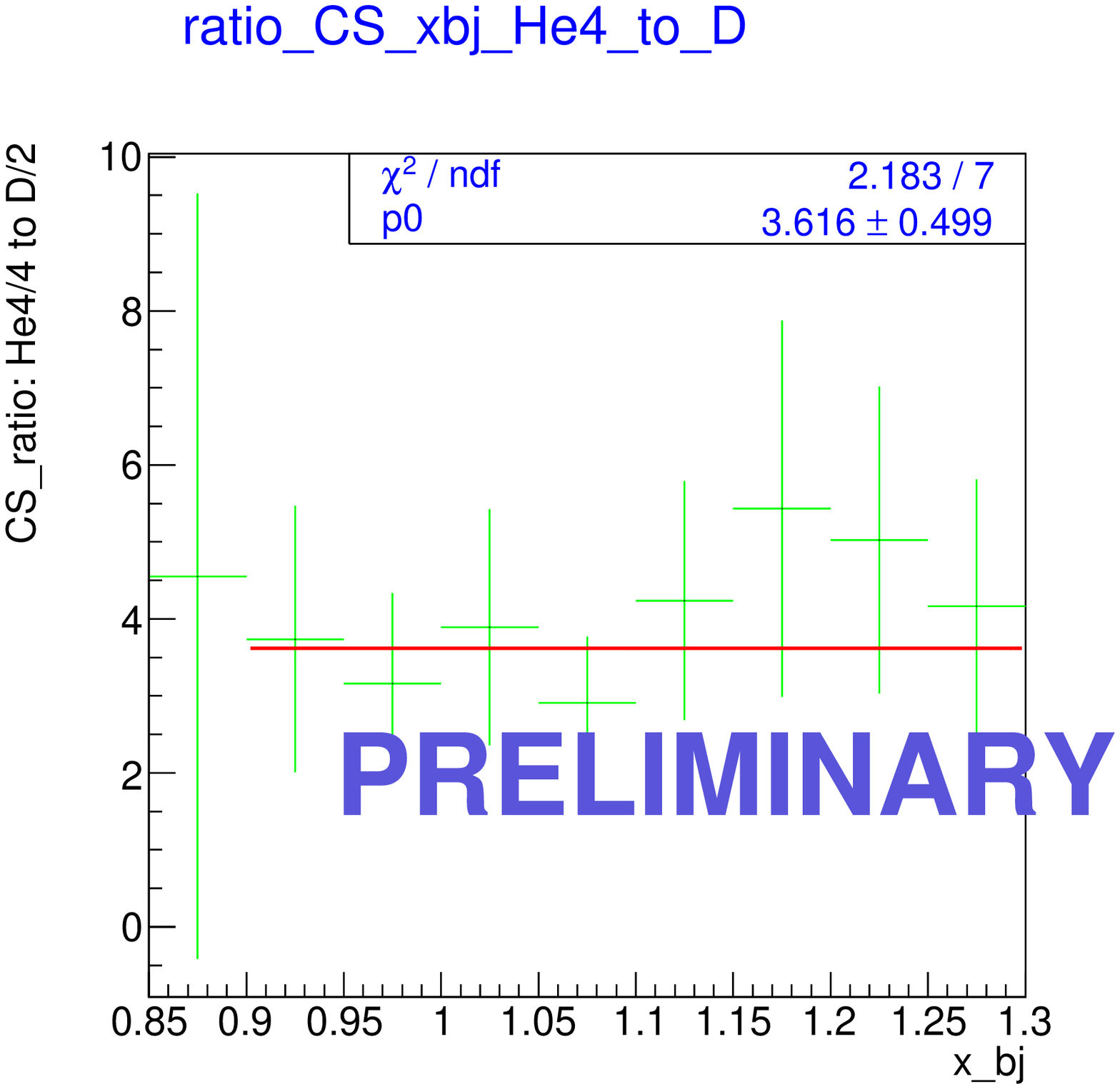} \\
  \includegraphics[width=\textwidth]{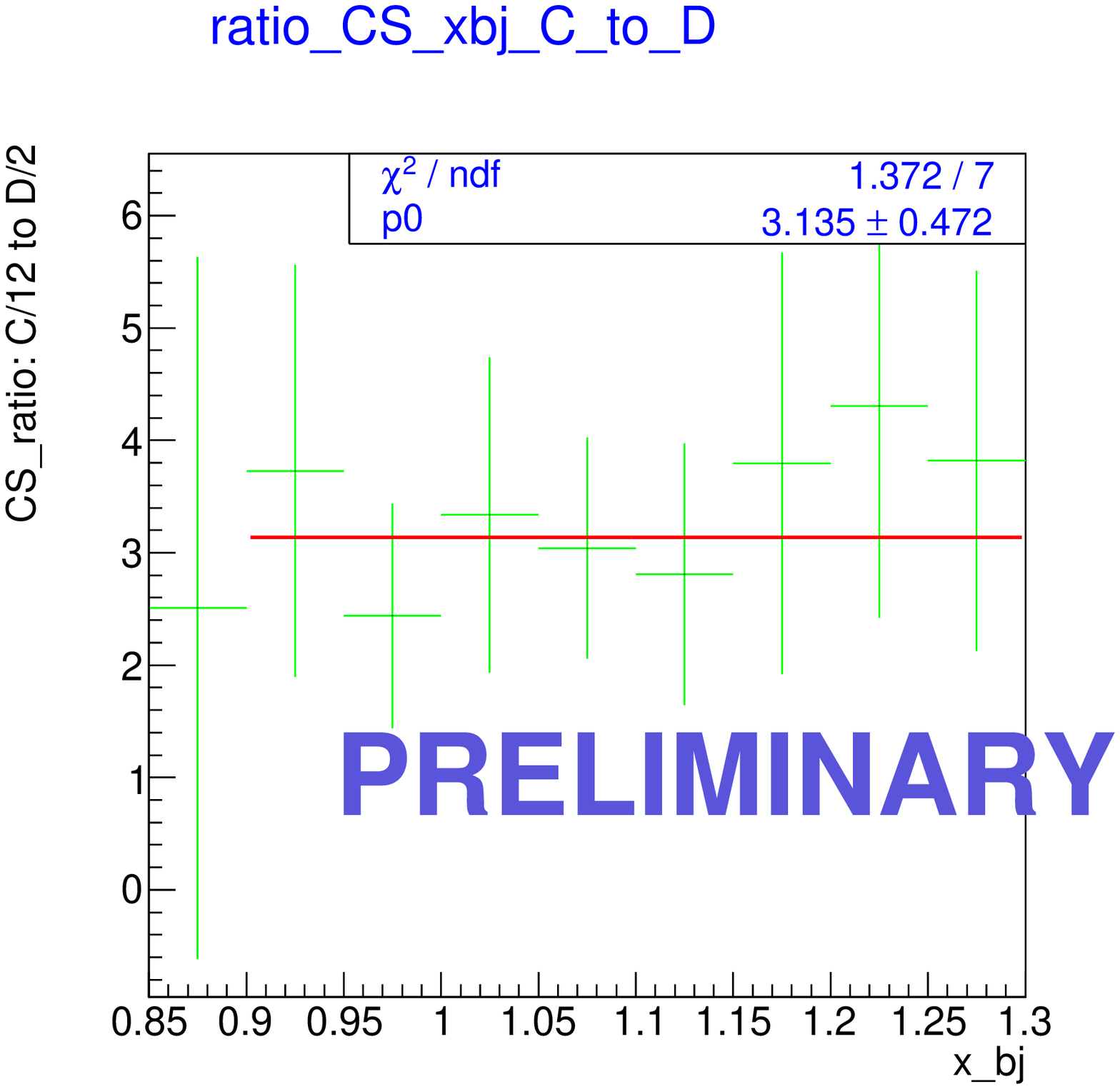} 
  \label{fig:e07006-ratios}
\end{tabular}
\end{minipage}
\hfill
\begin{minipage}[t]{0.3\textwidth}
\hrule height 0pt
\vskip -3mm
  \caption[E07006: Semi-Inclusive Ratios.]{Semi-inclusive cross-section ratios
    versus $x_{bj}$.  The top panel is the ratio of $^4$He to deuterium, and the 
    bottom panel is the ratio of $^{12}$C to deuterium.}
\end{minipage}
\end{center}
\end{figure}

In case of the semi-inclusive cross-section ratios of nuclei versus $x_{bj}$, 
we found a {\it flat} region starting around 0.9 up to 1.3, which begins below 
the flat region of the inclusive ratio ($x_{bj} >$~1.3), see Fig.~\ref{fig:e07006-ratios}.  
The ratio value is not equal to the $a_{2}$ value.  Further study is required to understand 
this phenomena. 

\clearpage





\subsection{E08009: $^4He(e,e'p)^3H$ Cross Sections up to $P_{miss}=0.632 GeV/c$}\label{sec:e08009}

\begin{center}
S. Iqbal, M. Ivanov, N. See \& SRC Collaboration, submitted by K. Aniol
\end{center}


\subsubsection{Introduction}
Experiments E07006 and E08009 ran in February, March and April of 2011. Data for kinematic settings
of 153 and 353 MeV/c for E08009 provide a thesis for Sophia Iqbal. In addition to these dedicated kinematic settings the Short Rangle Correlation(SRC) experiment also obtained data at kinematic settings
out to 632 MeV/c which could be analyzed for the two body break up channel p + triton. These higher
missing momenta data were collected using about 4 to 5 $\mu A$ current but sufficient accumulated charge
was measured to be able to extract cross sections beyond the original goal set for E08009. Moreover,
the large acceptances of the Hall A spectrometers allowed for cross sections to be determined across
a larger missing momentum range than the central value kinematic settings would suggest.

The electron spectrometer was fixed in angle and central momentum while the proton spectrometer's
angles and central momenta were changed. The electron arm settings are in table~\ref{tab:central_e}.

\begin{table}[hbt]
\begin{tabular}{|c|c|}
\hline
incident beam energy&   4.4506 GeV\\
electron spectrometer angle & 20.3$^\circ$ \\
electron spectrometer momentum & 3.602 GeV/c \\
Q$^2$ & 2.0 (GeV/c)$^2$ \\
Bjorken $x_b$ & 1.24\\
\hline
\end{tabular}
\caption{Electron spectrometer kinematic settings for E08009.}
\label{tab:central_e}
\end{table}

\subsubsection{Experimental cross sections}

Radiative corrections are obtained from the GEANT simulation of a p + triton final state
 missing energy spectrum at the spectrometer apertures.
Experimental cross sections are given in table~\ref{tab:xsect}.

\begin{table}[hbt]
\begin{tabular}{|c|c|c|c|c|}
\hline
$P_{miss}$ &  153 & 353 & 466 & 632  \\
(MeV/c)    &$\theta_p=47^\circ$ &$\theta_p=38.5^\circ$ &$\theta_p=33.5^\circ$ & $\theta_p=29^\circ$\\
\hline
25    &$(3.38\pm0.52)$&  &  &  \\
\hline
75    &$(1.13\pm0.17)$&  &  &  \\
\hline
125   &$(3.13\pm0.48)\times 10^{-1}$&  &  &  \\
\hline
175   &$(7.18\pm0.11)\times 10^{-2}$&  &  &  \\
\hline
225  &$(1.44\pm0.22)\times 10^{-2}$&$(4.40\pm0.14)\times 10^{-3}$  &  &  \\
\hline
275  &$(3.06\pm0.57)\times 10^{-3}$&$(1.27\pm0.03)\times 10^{-3}$&   &   \\
\hline
325  &    &$(6.11\pm0.14)\times 10^{-4}$ &  &  \\
\hline
375  &    &$(3.57\pm0.88)\times 10^{-4}$    &    &    \\
\hline
425  &    &$(1.44\pm0.59)\times 10^{-4}$    &$(6.59\pm2.7)\times 10^{-4}$    &    \\
\hline
475  &    &    &$(3.22\pm0.89)\times 10^{-4}$    &    \\
\hline
525  &    &    &$(1.68\pm0.45)\times 10^{-4}$    &    \\
\hline
575  &    &    &$(0.91\pm0.43)\times 10^{-4}$    &    \\
\hline
632  &    &    &    &$(3.7\pm2.3)\times 10^{-5}$    \\
\hline
\end{tabular}
\caption[E08009: Cross sections for $^4He(e,e'p)^3H$]{Cross sections for $^4He(e,e'p)^3H$ from E08009, for different
kinematical settings given by the proton spectrometer central angle. Units are $nb/sr^2/MeV$.}
\label{tab:xsect}
\end{table}

\subsubsection{Theoretical cross sections}

\begin{figure}[!ht]
\begin{center}
\begin{minipage}[t]{0.68\textwidth}
\hrule height 0pt
\includegraphics[width=\textwidth]{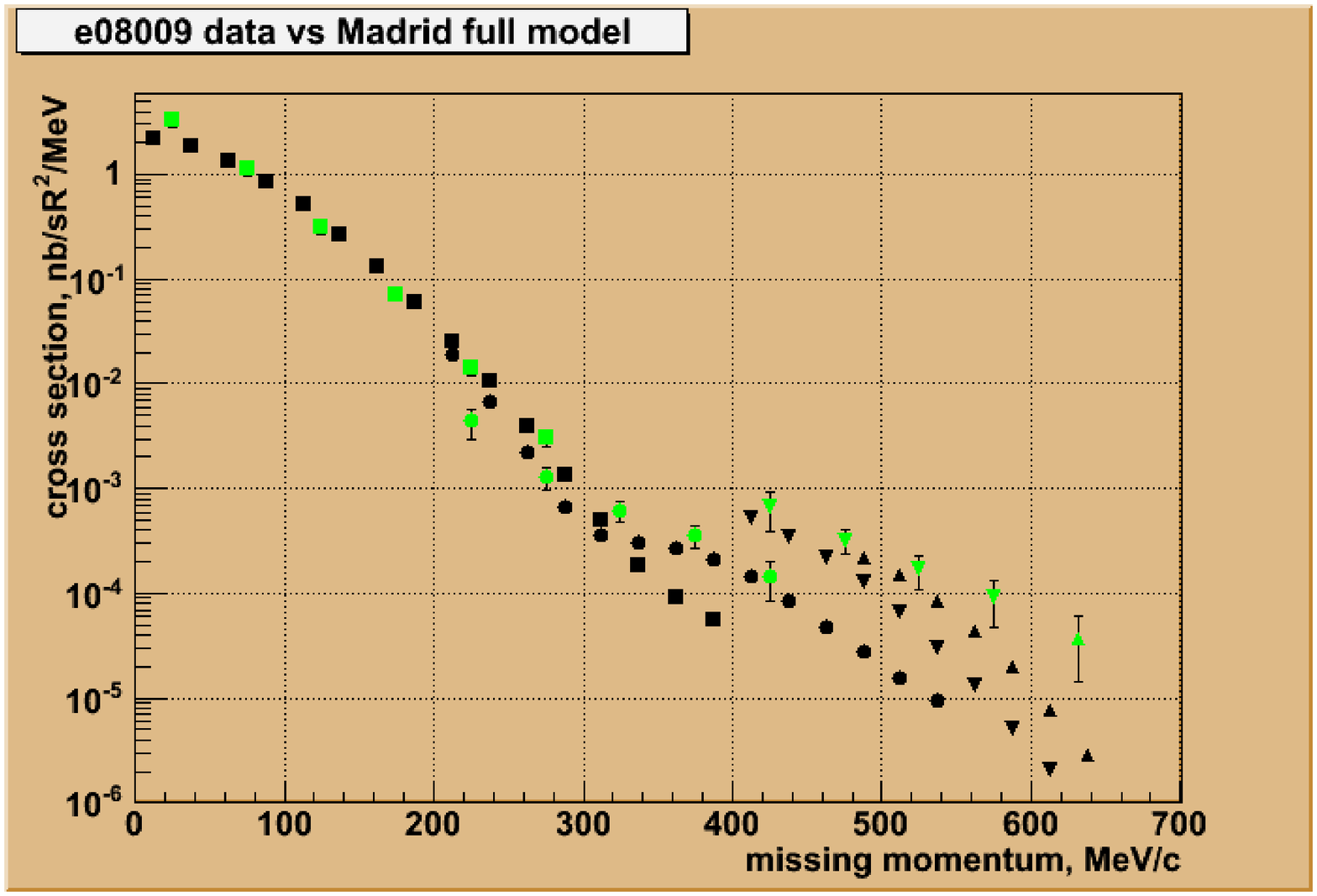}
\end{minipage}
\begin{minipage}[t]{0.3\textwidth}
\hrule height 0pt
\vskip -3mm
\caption[E08009: Data compared to full Madrid theoretical calculations.]{E08009 Data in green compared to full Madrid theoretical calculations in black. Squares
are for the 153 MeV/c setting, circles are for 353 MeV/c setting, inverted triangles are for
the 466 MeV/c setting and triangles are for the 632 MeV/c setting.}
\label{fig:data_full}
\end{minipage}
\end{center}
\end{figure}

\begin{figure}[!ht]
\begin{center}
\begin{minipage}[t]{0.68\textwidth}
\hrule height 0pt
\includegraphics[width=\textwidth]{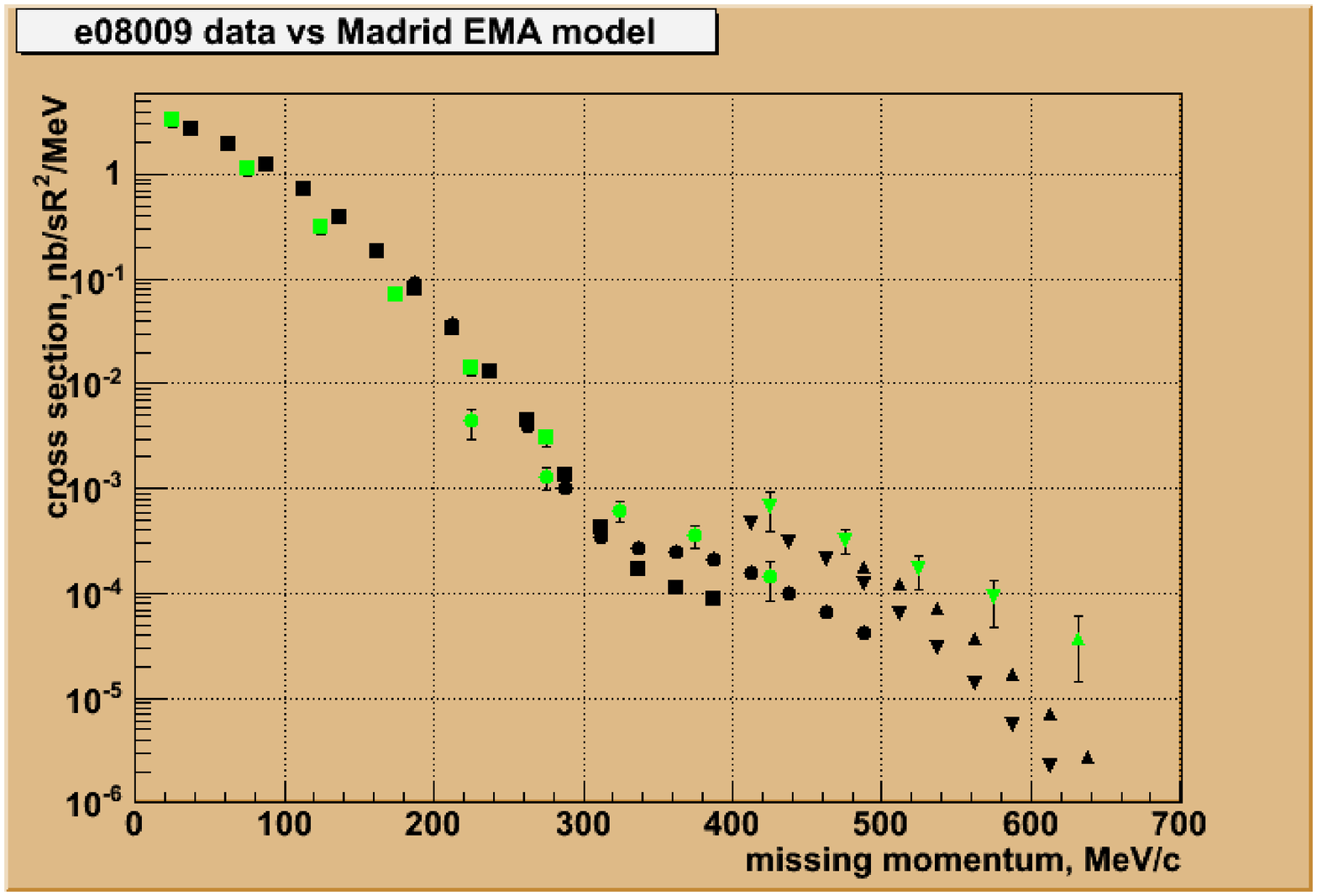}
\end{minipage}
\begin{minipage}[t]{0.3\textwidth}
\hrule height 0pt
\vskip -3mm
\caption[E08009: Data compared to full Madrid theoretical calculations.]{E08009 Data in green compared to EMA Madrid theoretical calculations in black. Squares
are for the 153 MeV/c setting, circles are for 353 MeV/c setting, inverted triangles are for
the 466 MeV/c setting and triangles are for the 632 MeV/c setting.}
\label{fig:data_EMA}
\end{minipage}
\end{center}
\end{figure}

Vertex values of the incident electron's momentum at various positions
within the target and the momenta of the scattered electron and
ejected proton were provide to the Madrid theory group for calculation of the cross section
at each event vertex in the GEANT simulation. The GEANT simulation also contains the
detected electron and proton momenta at the spectrometers' apertures. In this way the
vertex cross section can be associated with the missing momentum at the apertures. The
GEANT simulation includes external and internal bremsstrahlung.
Theoretical cross sections
 integrated over the experimental acceptances for
the full Madrid treatment and using the EMA treatment are in tables~\ref{tab:madridfull} and
~\ref{tab:madridEMA}. Plots of the data for the two theoretical treatments are shown in
figures~\ref{fig:data_full} and ~\ref{fig:data_EMA}.

\begin{table}[hbt]
\begin{tabular}{|c|c|c|c|c|}
\hline
$P_{miss}$ &  153 & 353 & 466 & 632  \\
(MeV/c)    &$\theta_p=47^\circ$ &$\theta_p=38.5^\circ$ &$\theta_p=33.5^\circ$ & $\theta_p=29^\circ$\\
\hline
12.5 &2.20585 &0 &0 &0 \\
\hline
37.5 &1.82871 &0 &0 &0 \\
\hline
62.5 &1.31389 &0 &0 &0 \\
\hline
87.5 &0.851553 &0 &0 &0 \\
\hline
112.5 &0.506994 &0 &0 &0 \\
\hline
137.5 &0.26989 &0 &0 &0 \\
\hline
162.5 &0.131086 &0 &0 &0 \\
\hline
187.5 &0.0598725 &0 &0 &0 \\
\hline
212.5 &0.0258303 &0.0191822 &0 &0 \\
\hline
237.5 &0.010439 &0.0067236 &0 &0 \\
\hline
262.5 &0.00395091 &0.00220872 &0 &0 \\
\hline
287.5 &0.00137024 &0.000668576 &0 &0 \\
\hline
312.5 &0.000490056 &0.000357781 &0 &0 \\
\hline
337.5 &0.000185816 &0.000309488 &0 &0 \\
\hline
362.5 &9.30929e-05 &0.00026867 &0 &0 \\
\hline
387.5 &5.63916e-05 &0.000207743 &0 &0 \\
\hline
412.5 &0 &0.000141879 &0.000528339 &0 \\
\hline
437.5 &0 &8.3657e-05 &0.000340153 &0 \\
\hline
462.5 &0 &4.80785e-05 &0.000222462 &0 \\
\hline
487.5 &0 &2.73925e-05 &0.000126155 &0.0002206 \\
\hline
512.5 &0 &1.54183e-05 &6.54197e-05 &0.0001491 \\
\hline
537.5 &0 &9.47828e-06 &2.97952e-05 &8.585e-05 \\
\hline
562.5 &0 &0 &1.28925e-05 &4.4e-05 \\
\hline
587.5 &0 &0 &5.07677e-06 &1.977e-05 \\
\hline
612.5 &0 &0 &2.00828e-06 &7.741e-06 \\
\hline
637.5 &0 &0 &8.3571e-07 &2.834e-06 \\
\hline
\end{tabular}
\caption[E08009: Theoretical cross sections for $^4He(e,e'p)^3H$ ]{Madrid full theoretical cross sections integrated over the
 experimental acceptances
 for $^4He(e,e'p)^3H$ for E08009, for different
kinematical settings given by the proton spectrometer central angle.
 Units are $nb/sr^2/MeV$.}
\label{tab:madridfull}
\end{table}

\begin{table}[hbt]
\begin{tabular}{|c|c|c|c|c|}
\hline
$P_{miss}$ &  153 & 353 & 466 & 632  \\
(MeV/c)    &$\theta_p=47^\circ$ &$\theta_p=38.5^\circ$ &$\theta_p=33.5^\circ$ & $\theta_p=29^\circ$\\
\hline
12.5 &0 &0 &0 &0 \\
\hline
37.5 &2.681 &0 &0 &0 \\
\hline
62.5 &1.916 &0 &0 &0 \\
\hline
87.5 &1.235 &0 &0 &0 \\
\hline
112.5 &0.729652 &0 &0 &0 \\
\hline
137.5 &0.383898 &0 &0 &0 \\
\hline
162.5 &0.183412 &0 &0 &0 \\
\hline
187.5 &0.0815901 &0.0903122 &0 &0 \\
\hline
212.5 &0.0338215 &0.0362774 &0 &0 \\
\hline
237.5 &0.0128213 &0.0129471 &0 &0 \\
\hline
262.5 &0.00443289 &0.0039332 &0 &0 \\
\hline
287.5 &0.00136237 &0.000998639 &0 &0 \\
\hline
312.5 &0.000431068 &0.000342315 &0 &0 \\
\hline
337.5 &0.000170451 &0.000264277 &0 &0 \\
\hline
362.5 &0.000112972 &0.00024869 &0 &0 \\
\hline
387.5 &8.81671e-05 &0.00020829 &0 &0 \\
\hline
412.5 &0 &0.000154708 &0.000455009 &0 \\
\hline
437.5 &0 &9.85333e-05 &0.000308199 &0 \\
\hline
462.5 &0 &6.48162e-05 &0.000206383 &0 \\
\hline
487.5 &0 &4.26083e-05 &0.000120555 &0.0001778 \\
\hline
512.5 &0 &0 &6.4348e-05 &0.0001215 \\
\hline
537.5 &0 &0 &3.03616e-05 &7.084e-05 \\
\hline
562.5 &0 &0 &1.35952e-05 &3.702e-05 \\
\hline
587.5 &0 &0 &5.52707e-06 &1.717e-05 \\
\hline
612.5 &0 &0 &2.25103e-06 &7.01e-06 \\
\hline
637.5 &0 &0 &9.48271e-07 &2.695e-06 \\
\hline
\end{tabular}
\caption{Madrid EMA theoretical cross sections integrated over the experimental acceptances for $^4He(e,e'p)^3H$ for E08009, for different
kinematical settings given by the proton spectrometer central angle. Units are $nb/sr^2/MeV$.}
\label{tab:madridEMA}
\end{table}

\clearpage
\newpage








\subsection{E08027: A Measurement of G$_2^P$ and the
  longitudinal-transverse spin polarizability}

\begin{center}
contributed by P. Zhu (USTC) for the E08-027 collaboration.
\end{center}

\subsubsection{Motivation}

Nucleon structure is described by the structure functions extracted
from inclusive cross sections. In unpolarized case, there are two
structure functions, $F_{1}$ and $F_{2}$. If the beam and target
are polarized, there are two additional spin-dependent structure functions,
$g_{1}$ and $g_{2}$. The $g_{1}$ structure function represents
the charge-weighted quark helicity distributions at the Bjorken limit.
However, $g_{2}$ has no simple interpretation in the naive parton
model. It is related to the higher-twist effects, i.e., quark-gluon
correlations. Measurements of the spin structure function (SSF) $g_{2}$
for the proton at low $Q^{2}$ are lacking. Currently the lowest momentum
transfer investigated is $1.3\, GeV^{2}$ by the RSS collaboration
\cite{PhysRevLett.98.132003}. 

The goal of this experiment is to measure the $g_{2}$ structure function
for the proton at low $Q^{2}.$ A measurement of the longitudinally-transverse
spin polarizability ($\delta_{LT}$) is expected to be a good test
of Chiral Perturbation Theory ($\chi PT$) since it is insensitive
to the $\pi-\Delta$ contribution, which is usually the main high-order
correction \cite{PhysRevD.87.054032}. The recent $\delta_{LT}$ data
for the neutron indicates a significant disagreement with the $\chi PT$
calculations \cite{PhysRevLett.93.152301}. However, the $\delta_{LT}$
data for the proton at low $Q^{2}$ does not exist yet. The $g_{2}$
data can provide a test of the Burkhardt-Cottingham sum rule. The
low $Q^{2}$ $g_{2}$ data will help improve the precision in the
hyperfine splitting of the hydrogen ground state, since the leading
uncertainty in the measurement of the hyperfine splitting in the hydrogen
ground state comes from the proton structure correction \cite{PhysRevLett.96.163001}.
The data from this experiment may also help to improve the precision
of the measurement for the proton charge radius.

\subsubsection{The Experiment}

\begin{figure}[h]
\begin{centering}
\includegraphics[width=0.35\paperwidth]{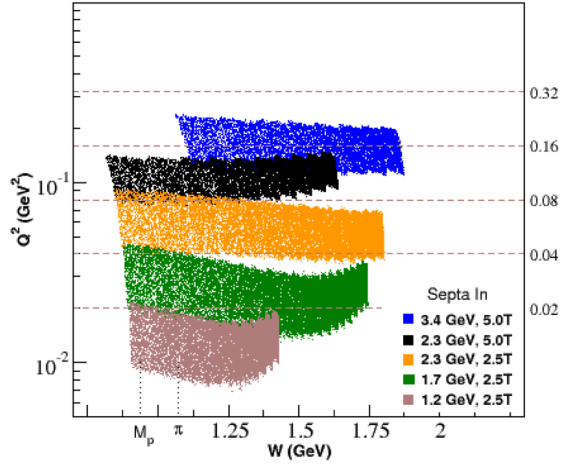} 
\par\end{centering}

\protect\protect\caption[E08027:  Kinematic coverage during the
experimental run period.]{\label{fig:Kinematic-coverage-during}Kinematic coverage during the
experimental run period. The numbers next to the vertical axis on
the right side are the constant $Q^{2}$ values where the moments
of the $g_{2}$ will be extracted.}
\end{figure}

The experiment successfully collected data from March to May, 2012.
A measurement of the scattered electrons in the reaction $\overrightarrow{p}(\overrightarrow{e},e')X$
at a scattering-angle of $5.69^{\circ}$ in the low $Q^{2}$ region
of $0.02<Q^{2}<0.2\, GeV^{2}$ was performed to obtain the proton
spin-dependent cross sections (see figure \ref{fig:Kinematic-coverage-during}).

\begin{figure}[h]
\begin{centering}
\includegraphics[width=0.8\linewidth]{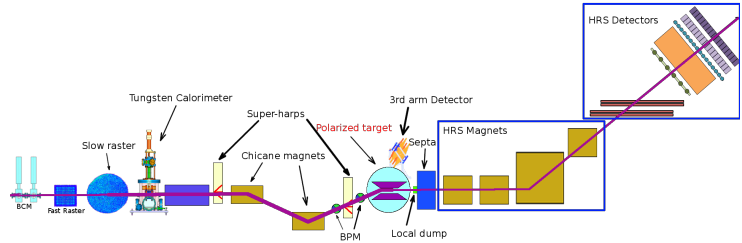} 
\par\end{centering}

\protect\protect\caption[E08027:  Installation of the $g_{2}^{p}$ experiment.]{\label{fig:Installation}Installation of the $g_{2}^{p}$ experiment. }
\end{figure}

A polarized $NH_{3}$ target operating at 1 K was used as the proton
target. The Dynamic Nuclear Polarization(DNP) process was used to
polarize the solid $NH_{3}$ target. To avoid too much depolarization
of the target, beam current was limited to $50-100$ nA during the
experiment. Since the existing beam current monitors (BCMs), beam
position monitors (BPMs) and calibration methods did not work at such
a low current range, new BPM and BCM receivers were designed and used
for low current condition. A pair of super-harps and a tungsten calorimeter
were installed to calibrate the BPMs and BCMs. To compensate for the
effect of the 2.5/5 T transverse target magnet field, two chicane
dipole magnets were installed. A pair of slow rasters were installed
for the first time in Hall A to spread the beam to a diameter of 2
cm, combining with a pair of fast rasters. To allow detection of scattered
elections at a $5.69^{\circ}$ scattering angle, the target was
installed at 876.93mm upstream from the pivot and a pair of septum
magnets were installed. A new scintillator detector was developed
and placed near the target to monitor the polarization of the beam
and target. The instruments used in the experiment are shown in figure
\ref{fig:Installation}. 

\subsubsection{Experimental Progress}

The target polarization was measured by the method of Nuclear Magnetic
Resonance (NMR). Thermal equilibrium measurements were used for calibrating
the readout from NMR. The average measured polarization for the 5T
magnet field is 70\%, while for the 2.5T magnetic field is 15\% .
A NIM paper was published for the $g_{2}^{p}$ target system \cite{Pierce201454}.

The standard HRS detectors had a good performance during the experiment.
We obtained higher than 99\% efficiency for the spectrumeter detectors
during experiment, including track efficiency with carefully examined
multitrack-events, Cherencov efficiency, lead glass calorimeter efficiency,
and S1 and S2m trigger scintillator efficiency. Also the pion contamination
was contained to below 0.4\% with the particle identification cuts.

The beam position and angle are used for calculating the scattering
angle and optimizing the optics result. Pedestals of the BPMs varied
significantly during the experiment. A data base was set up to track
the variation of the pedestals with time. A 2-Hz software filter was
used to improve the signal/noise ratio. A new method to analyze the
data of BPMs and harps, which was optimized for low beam current and
large raster size, was developed. In order to transport the position
from two BPMs to the target, several transport functions are fitted
by using the simulation with the known target-magnetic field map.

Because of the low pass filter used in the BPM receiver and the time
delay of the BPM readout, the BPM information is not enough to get
the beam position event-by-event. To get the position and angle event-by-event,
the raster magnet current information is used to account for the beam
motion caused by the rasters. A calibrated conversion factor is needed
to convert the raster current to beam position shift. For the slow
raster, a carbon hole was used to convert the raster current to the
size of the beam spot, while for the fast raster, the beam spot size
at the calibrated BPMs was used. Since the distance of two BPMs is
only 26.5 cm, while the distance of the 2nd BPM to the target is 69
cm, the uncertainty of the beam position at the target is magnified
by a factor of 5. The uncertainty at the target is 1-2 mm for position
and 1-2 mrad for angle.

A new Monte-Carlo simulation package was developed to study the spectrometer
acceptance and the optics calibration with the target field. It has
been tuned to work with the affect of the target and septum fields.
The package was developed with an optimized Runge-Kutta method with
self-adjusting step length to improve the speed and accuracy, based
on the hall A Single Arm Monte-Carlo (SAMC) package. Several different
cross-section models and energy-loss models are included in the elastic
and resonance kinematic regions. The simulation results are also used
to compare with the results in the packing fraction study.

The purpose of the HRS optics study is to reconstruct the kinematic
variables of the scattered electrons at the reaction point. Currently,
the optics data with no target field has been optimized for both the
left and right HRS, which allows us to remove the additional complexity
of the target field and focus on the septa and HRS magnets. The $\theta$,
$\phi$ and $\delta$ related optics matrix elements are calibrated
with the standard sieve slit method \cite{nilangaoptics}. At the
very forward scattering angle of $5.69^{\circ}$, foil targets with
a large $z$ position separation are needed to calibrate the $y_{tg}$
related optics matrix elements. A single foil carbon target and the
aluminum entrance window of the target chamber are used for this purpose.
This will make the resolutions slightly worse but still satisfy our
requirement. The resolutions are close to the nominal performance
of the HRS system as shown in table \ref{tab:Performance-optics}.
\begin{table}[h]
\begin{centering}
\begin{tabular}{|c|c|c|c|}
\hline 
RMS & LHRS & RHRS & Nominal performance \cite{nilangaoptics}\tabularnewline
\hline 
\hline 
$\delta${[}dp{]}  & $1.5\times10^{-4}$  & $2.4\times10^{-4}$ & $1.1\times10^{-4}$\tabularnewline
\hline 
$\theta${[}out of plane angle{]}  & 1.59 mrad  & 1.57 mrad & 2.55 mrad\tabularnewline
\hline 
y  & 3.3 mm  & 2.9 mm & 1.7 mm\tabularnewline
\hline 
$\phi${[}in plane angle{]}  & 0.99 mrad  & 0.82 mrad & 0.85 mrad\tabularnewline
\hline 
\end{tabular}
\par\end{centering}

\protect\protect\caption{\label{tab:Performance-optics}Performance summary of RMS values for
optics study without target field}
\end{table}

In $g_{2}^{p}$ setting, the strong transverse target field makes
the optics study more challenging. To deal with this target field,
the reconstruction process is separated into two parts. The first
part, containing the septum magnet and HRS, is assumed to be represented
by the matrix with no target field which has been described above.
Unfortunately, the configuration changes during the experiment because
of the broken septum magnet, which requires the matrix elements to
be re-calibrated. The simulation package mentioned above is used to
calculate the reference angles of the fits for the recalibration.
The second part, the target field region, is treated only with a ray-tracing
method. The same simulation package is also used here to calculate
the trajectory of the scattered electrons. The calibration has been
completed for LHRS and will be done soon for RHRS.

In additional to the calibration of the optics matrix, the central
angle of the spectrometer system is studied with two different methods:
using survey information or using double elastic peaks \cite{kiadoptics}.
The survey information provides smaller systematic uncertainty and
is the one being used. The central angle, together with the relative
scattering angle reconstructed by the optics matrix, is used to calculate
the scattering angle of the out-going electrons. 

\begin{figure}[h]
\begin{centering}
\includegraphics[width=0.7\linewidth]{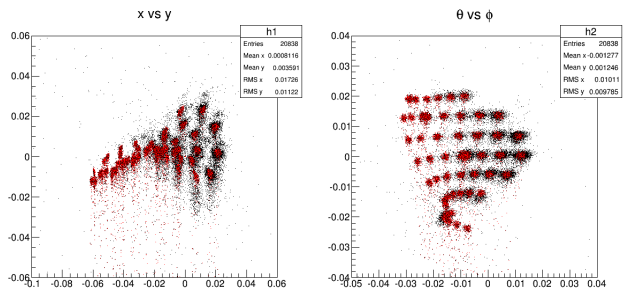}
\par\end{centering}

\protect\caption[E08027:  Match in focal plane for acceptance study]{\label{fig:Match-in-focal}Match in focal plane for acceptance study}

\end{figure}

In addition, an acceptance study is underway. A correction was made
to the SNAKE model to match data at focal plane well in good septum
situation (figure \ref{fig:Match-in-focal}).
 With this updated SNAKE model, forward and reverse transport functions
have been fitted to describe the electron trajectories from the target
plane to the focal plane (without target field). These functions have
been incorporated into the simulation package. The acceptance for
the other septum configurations will be studied soon. 

The $NH_{3}$ target cell is comprised of ammonia beads and liquid
helium. The packing fraction, or the ratio of the length of ammonia
to the total target length, must be understood for dilution analysis.
To extract the packing fraction, the elastic yields are compared from
a production run and a ``dummy'' run, where the target cell is filled
only with liquid helium. Currently, the fitting routine is being optimized
to remove contamination to the elastic peak from the quasi-elastic
peak. The packing fraction from one of our setting is 0.6.

\begin{figure}[h]
\begin{centering}
\includegraphics[width=0.5\linewidth]{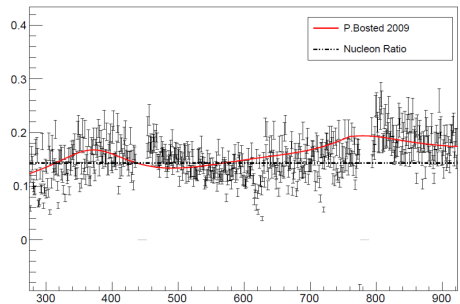}
\par\end{centering}

\protect\caption[E08027: Dilution factor for the 3.35GeV beam energy]{\label{fig:Dilution-factor}Dilution factor for the 3.35GeV beam energy,
5T transverse field setting; x axis is $\nu(MeV)$}

\end{figure}

The measured asymmetry is diluted by contributions from the nitrogen
in the ammonia target material, helium used to cool the target, and
the aluminum target end caps. This contamination is removed, giving
us a true proton asymmetry, using a dilution factor analysis, which
accounts for scattering from the unpolarized material. The aluminum
and helium background is determined from experimental data, but the
nitrogen background is more complicated since we do not have pure
nitrogen data (we only took data on a carbon target). Since the Small
Angle GDH experiment has similar kinematics to g2p, we can use the
saGDH nitrogen data set to tune the Bosted model for use at the g2p
kinematics. Elastic and inelastic radiative corrections have been
completed on the saGDH nitrogen data and the Bosted model has been
tuned to $\pm5\%$ level \cite{ryanrcsagdh}. The actual nitrogen
background is scaled by using the tuned Bosted model with the g2p
carbon data. Figure \ref{fig:Dilution-factor} shows the dilution
factor for one of our kinematic settings.

Once the acceptance study, the packing fraction and the dilution studies
are complete, the physics asymmetries and the cross sections can be
extracted. Preliminary results are expected in another year or so.

\bibliographystyle{elsarticle-num}
\bibliography{e08027-2014}


\clearpage

\section{Publications}

\newcommand{\etal}{{\em et~al.}}

Publications published during 2014, either in preprint or finally, based
on experiments run in Hall A of  Jefferson Lab.

\begin{enumerate}

\item{C J Horowitz \etal, 
A way forward in the study of the symmetry energy: experiment, theory, and observation,
\href{http://iopscience.iop.org/0954-3899/41/9/093001/}{J. Phys. G: Nucl. Part. Phys. 41 (2014) 093001}
}

\item{S. Golge, B. Vlahovic, B. Wojtsekhowski, 
High-intensity positron microprobe at the Thomas Jefferson National Accelerator Facility,
\href{http://scitation.aip.org/content/aip/journal/jap/115/23/10.1063/1.4884781}{J. Appl. Phys. 115, 234907 (2014)}
}

\item{A. Camsonne \etal, 
JLab Measurement of the $^4$He Charge Form Factor at Large Momentum Transfers,
\href{http://journals.aps.org/prl/abstract/10.1103/PhysRevLett.112.132503}{Phys. Rev. Lett. 112, 132503 (2014)}
}

\item{M. Mihovilovi\u{c} \etal, 
Measurement of Double-Polarization Asymmetries in the Quasielastic $^3$He(e,e′d) Process,
\href{http://journals.aps.org/prl/abstract/10.1103/PhysRevLett.113.232505}{Phys. Rev. Lett. 113, 232505 (2014)}
}

\item{The Jefferson Lab PVDIS Collaboration, 
Measurement of parity violation in electron–quark scattering,
\href{http://www.nature.com/nature/journal/v506/n7486/full/nature12964.html}{Nature 506, 67–70 (06 February 2014)}
}

\item{Y. Zhang \etal, 
Measurement of “pretzelosity” asymmetry of charged pion production in semi-inclusive deep inelastic scattering on a polarized He3 target,
\href{http://journals.aps.org/prc/abstract/10.1103/PhysRevC.90.055209}{Phys. Rev. C 90, 055209 (2014)}
}

\item{P Monaghan \etal, 
Measurement of the $^{12}$C(e,e′p)11B two-body breakup reaction at high missing momentum,
\href{http://iopscience.iop.org/0954-3899/41/10/105109/}{J. Phys. G: Nucl. Part. Phys. 41 (2014) 105109}
}

\item{J. Katich \etal, 
Measurement of the Target-Normal Single-Spin Asymmetry in Deep-Inelastic Scattering from the Reaction $^3$He$^\uparrow$(e,e′)X,
\href{http://journals.aps.org/prl/abstract/10.1103/PhysRevLett.113.022502}{Phys. Rev. Lett. 113, 022502 (2014)}
}

\item{O. Hen \etal, 
Momentum sharing in imbalanced Fermi systems,
\href{http://www.sciencemag.org/content/346/6209/614}{Science 346, 614 (2014)}
}

\item{M. Posik \etal, 
Precision Measurement of the Neutron Twist-3 Matrix Element d$^n_2$: Probing Color Forces,
\href{http://journals.aps.org/prl/abstract/10.1103/PhysRevLett.113.022002}{Phys. Rev. Lett. 113, 022002}
}

\item{Igor Korover \etal, 
Probing the Repulsive Core of the Nucleon-Nucleon Interaction via the $^4$He(e,e′pN) Triple-Coincidence Reaction,
\href{http://journals.aps.org/prl/abstract/10.1103/PhysRevLett.113.022501}{Phys. Rev. Lett. 113, 022501 (2014)}
}

\item{Y. X. Zhao \etal, 
Single Spin Asymmetries in Charged Kaon Production from Semi-Inclusive Deep Inelastic Scattering on a Transversely Polarized $^3$He Target,
\href{http://journals.aps.org/prc/abstract/10.1103/PhysRevC.90.055201}{Phys. Rev. C.90.055201, 2014}
}

\item{K. Allada  \etal, 
Single spin asymmetries of inclusive hadrons produced in electron scattering from a transversely polarized $^3$He target,
\href{http://journals.aps.org/prc/abstract/10.1103/PhysRevC.89.042201}{Phys. Rev. C 89, 042201(R), 2014}
}

\item{Simona Malace, David Gaskell, Douglas W. Higinbotham, Ian Cloet, 
The Challenge of the EMC Effect: existing data and future directions,
\href{http://www.worldscientific.com/doi/abs/10.1142/S0218301314300136}{Int. J. Mod. Phys. E 23 (2014) 1430013}
}

\item{Krishna S. Kumar \etal, The MOLLER Experiment: An Ultra-Precise
    Measurement of the Weak Mixing Angle using M{\o}ller Scattering,
\href{http://arxiv.org/abs/1411.4088}{arxiv:1411.4088}
}

\item{B. Wojtsekhowski \etal, 
On measurement of the isotropy of the speed of light,
\href{http://arxiv.org/abs/1409.6373}{arxiv:1409.6373}
}

\item{Douglas W. Higinbotham, Or Hen, 
Comment on "Measurement of 2- and 3-nucleon short range correlation probabilities in nuclei",
\href{http://arxiv.org/abs/1409.3069}{arxiv:1409.3069}
}

\item{D. S. Parno \etal, 
Precision Measurements of A$^n_1$ in the Deep Inelastic Regime,
\href{http://arxiv.org/abs/1406.1207}{arxiv:1406.1207}
}

\end{enumerate}


\section{Theses}
\begin{enumerate}

\item{\emph{Measurement of the photon electroproduction cross section at JLAB with the goal of performing a Rosenbluth separation of the DVCS contribution}}\\
Alejandro Marti Jimenez-Arguello\\
\url{https://misportal.jlab.org/ul/publications/view_pub.cfm?pub_id=13562}

\item{\emph{A Precision Measurement of the Neutron D2: Probing the
      Color Force}}\\
Matthew Posik\\
\url{https://misportal.jlab.org/ul/publications/view_pub.cfm?pub_id=13193}

\end{enumerate}

\clearpage

\section{Hall A Collaboration Member List, 2014}
\begin{multicols}{3}

\RaggedRight

{\parindent 0cm 
{\bf Argonne National Lab}\\
John Arrington\\
Paul Reimer\\
Xiaohui Zhan\\

\bigskip{\bf Brookhaven National Lab}\\
Xin Qian\\

\bigskip{\bf Budker Institute of Nuclear Physics}\\
Dima Nikolenko\\
Igor Rachek\\

\bigskip{\bf Cairo University}\\
Hassan Ibrahim\\


\bigskip{\bf California State University}\\
Konrad A. Aniol\\
Martin B. Epstein\\
Dimitri Margaziotis\\

\bigskip{\bf Carnegie Mellon University}\\
Gregg Franklin\\
Vahe Mamyan\\
Brian Quinn\\

\bigskip{\bf The Catholic University of America}\\
Marco Carmignotto\\
Tanja Horn\\ 
Indra Sapkota\\

\bigskip{\bf Commissariat a l’Energie Atomique - Saclay}\\
Maxime Defurne \\
Nicole d'Hose\\
Eric Fuchey\\
Franck Sabatie\\

\bigskip{\bf China Institute of Atomic Energy (CIAE)}\\
Xiaomei Li\\
Shuhua Zhou\\

\bigskip{\bf Christopher Newport University}\\
Ed Brash\\

\bigskip{\bf College of William and Mary}\\
David S. Armstrong\\
Carlos Ayerbe Gayoso\\
Todd Averett\\
Juan Carlos Cornejo\\
Melissa Cummings\\
Wouter Deconinck\\
 Keith Griffioen\\
Joe Katich\\
Charles Perdrisat\\
Yang Wang\\ 
Huan Yao\\
Bo Zhao\\


\bigskip{\bf Duquesne University}\\
Fatiha Benmokhtar\\

\bigskip{\bf Duke University}\\
Steve Churchwell\\
Haiyan Gao\\
Calvin Howell\\
Min Huang\\
Richard Walter\\
Qiujian Ye\\

\bigskip{\bf Faculte des Sciences de Monastir (Tunisia)}\\
Malek Mazouz\\

\bigskip{\bf Florida International University}\\
Armando Acha\\
Werner Boeglin\\
Luminiya Coman\\
Marius Coman\\
Lei Guo\\
Hari Khanal\\
Laird Kramer\\
Pete Markowitz \\
Brian Raue\\
Jeorg Reinhold\\

\bigskip{\bf Forschungszentrum Rossendorf Institut f\"ur Kern- und. Hadronenphysik}\\
Frank Dohrmann \\


\bigskip{\bf Gesellschaft fur Schwerionenforschung (GSI)}\\
Javier Rodriguez Vignote\\

\bigskip{\bf Hampton University}\\
Eric Christy\\
Leon Cole\\
Peter Monaghan\\

\bigskip{\bf Harvard University}\\
Richard Wilson\\

\bigskip{\bf Hebrew University of Jerusalem}\\
Moshe Friedman\\
Aidan Kelleher\\
Guy Ron \\

\bigskip{\bf Huangshan University}\\
Hai-jiang Lu\\
XinHu Yan\\

\bigskip{\bf Idaho State University}\\
Mahbub Khandaker\\
Dustin McNulty\\

\bigskip{\bf INFN/Bari}\\
Raffaele de Leo\\

\bigskip{\bf INFN/Catania}\\
Antonio Guisa\\
Francesco Mammolit\\
Giuseppe Russo\\
Concetta Maria Sutera\\

\bigskip{\bf INFN/Lecce}\\
Roberto Perrino\\

\bigskip{\bf INFN/Roma}\\
Marco Capogni\\
Evaristo Cisbani\\
Francesco Cusanno\\
Fulvio De Persio\\
Alessio Del Dotto\\
Cristiano Fanelli\\
Salvatore Frullani\\
Franco Garibaldi\\
Franco Meddi\\
Guido Maria Urciuoli\\

\bigskip{\bf Institute of Modern Physics, Chinese Academy of Sciences}\\
Xurong Chen\\

\bigskip{\bf Institut de Physique Nucleaire - Orsay}\\
Camille Desnault\\
Alejandro Marti Jimenez-Arguello\\
Carlos Munoz Camacho\\
Rafayel Paremuzyan\\

\bigskip{\bf ISN Grenoble}\\
Eric Voutier\\

\bigskip{\bf James Madison University}\\
Gabriel Niculescu\\
Ioana Niculescu\\

\bigskip{\bf Jefferson Lab}\\
Alexandre Camsonne\\
Larry Cardman\\
Jian-Ping Chen\\
Eugene Chudakov\\
Mark Dalton\\
Kees de Jager\\
Alexandre Deur\\
Ed Folts\\
David Gaskell\\
Javier Gomez\\
Ole Hansen\\
Douglas Higinbotham\\
Mark K. Jones\\
Thia Keppel\\
John Lerose\\
Simona Malace\\
Bert Manzlak\\
David Meekins\\
Robert Michaels\\
Bryan Moffit\\
Sirish Nanda\\
Noel Okay\\
Lubomir Pentchev\\
Yi Qiang\\
Lester Richardson\\
Yves Roblin\\
Brad Sawatzky\\
Jack Segal\\
Dennis Skopik\\
Patricia Solvignon\\
Mark Stevens\\
Riad Suleiman\\
Stephanie Tysor\\
Bogdan Wojtsekowski\\
Jixie Zhang\\

\bigskip{\bf Jozef Stefan Institute}\\
Miha Mihovilovic\\
Simon Sirca\\

\bigskip{\bf Kent State University}\\
Bryon Anderson\\
Mina Katramatou\\
Elena Khrosinkova\\
Richard Madey\\
Mark Manley\\
Gerassimos G. Petratos\\
Larry Selvey\\
John Watson\\

\bigskip{\bf Kharkov Institute of Physics and Technology}\\
Oleksandr Glamazdin\\
Viktor Gorbenko\\
Roman Pomatsalyuk\\
Vadym Vereshchaka\\

\bigskip{\bf Kharkov State University}\\
Pavel Sorokin\\

\bigskip{\bf Khalifa University}\\
Issam Qattan\\

\bigskip{\bf Lanzhou University}\\
Bitao Hu\\
Yi Zhang\\

\bigskip{\bf Longwood University}\\
Tim Holmstrom\\
Keith Rider\\
Jeremy St. John\\
Vincent Sulkosky\\
Wolfgang Troth\\

\bigskip{\bf Los Alamos Laboratory}\\
Jin Huang\\
Xiaodong Jiang\\
Ming Xiong Liu\\

\bigskip{\bf LPC Clermont-Ferrand France}\\
Pierre Bertin\\
Helene Fonvielle\\

\bigskip{\bf Mississippi State University}\\
Dipangkar Dutta\\
Mitra Shabestari\\
Amrendra Narayan\\
Nuruzzaman\\

\bigskip{\bf Massachusetts Institute of Technology}\\
Kalyan Allada\\
Bill Bertozzi\\
Shalev Gilad\\
Navaphon ``Tai'' Muangma\\
Kai Pan\\
Cesar Fernandez Ramirez\\
Rupesh Silwal\\

\bigskip{\bf Mountain View Collage}\\
Ramesh Subedi\\

\bigskip{\bf Negev Nuclear Research Center}\\
Arie Beck\\
Sharon Beck\\

\bigskip{\bf NIKHEF}\\
Jeff Templon\\

\bigskip{\bf Norfolk State University}\\
Wendy Hinton\\
Vina Punjabi\\

\bigskip{\bf North Carolina Central University}\\
Benjamin Crowe\\
Branislav (Branko) Vlahovic\\

\bigskip{\bf North Carolina A\&T State University}\\
Ashot Gasparian\\

\bigskip{\bf Northwestern University}\\
Ralph Segel\\

\bigskip{\bf Ohio University}\\
Mongi Dlamini\\
Norman Israel\\
Paul King\\
Julie Roche\\

\bigskip{\bf Old Dominion University}\\
S. Lee Allison\\
Gagik Gavalian\\
Mohamed Hafez\\
Charles Hyde\\
Kijun Park\\
Hashir Rashad \\
Larry Weinstein\\

\bigskip{\bf Peterburg Nuclear Physics Institute}\\
Viacheslav Kuznetsov \\

\bigskip{\bf Regina University}\\
Alexander Kozlov\\
Andrei Semenov\\

\bigskip{\bf Rutgers University}\\
Lamiaa El Fassi\\
Ron Gilman\\
Gerfried Kumbartzki\\
Katherine Myers\\
Ronald Ransome\\
Yawei Zhang\\

\bigskip{\bf Saint Norbert College}\\
Michael Olson\\

\bigskip{\bf Seoul National University}\\
Seonho Choi\\
Byungwuek Lee\\

\bigskip{\bf Smith College}\\
Piotr Decowski\\

\bigskip{\bf St Mary's University}\\
Davis Anez\\
Adam Sarty\\


\bigskip{\bf Stony Brook University}\\
Rouven Essig\\

\bigskip{\bf Syracuse University}\\
Zafar Ahmed\\
Richard Holmes\\
Paul A. Souder\\

\bigskip{\bf Technische Universit\"at M\"unchen}\\
Jaideep Singh\\

\bigskip{\bf Tel Aviv University}\\
Nathaniel Bubis\\
Or Chen\\
Igor Korover\\
Jechiel Lichtenstadt\\
Eli Piasetzky\\
Ishay Pomerantz\\
Ran Shneor\\
Israel Yaron\\

\bigskip{\bf Temple University}\\
David Flay\\
Zein-Eddine Meziani\\
Michael Paolone\\
Matthew Posik\\
Nikos Sparveris\\

\bigskip{\bf Tohoku University}\\
Kouichi Kino\\
Kazushige Maeda\\
Teijiro Saito\\
Tatsuo Terasawa\\
H. Tsubota\\

\bigskip{\bf Tsinghua University}\\
Zhigang Xiao\\

\bigskip{\bf Universidad Complutense de Madrid (UCM)}\\
Joaquin Lopez Herraiz\\
Luis Mario Fraile\\
Maria Christina Martinez Perez\\
Jose  Udias Moinelo\\

\bigskip{\bf University of Connecticut}\\
Andrew Puckett\\

\bigskip{\bf Universitat Pavia}\\
Sigfrido Boffi\\

\bigskip{\bf University ``La Sapienza'' of Rome}\\
Cristiano Fanelli \\
Fulvio De Persio\\

\bigskip{\bf University of Glasgow}\\
John Annand\\
David Hamilton\\
Dave Ireland\\
Ken Livingston\\
Dan Protopopescu\\
Guenther Rosner\\
Johan Sjoegren\\

\bigskip{\bf University of Illinois}\\
Ting Chang\\
Areg Danagoulian\\
J.C. Peng\\
Mike Roedelbronn\\
Youcai Wang\\
Lindgyan Zhu\\

\bigskip{\bf University of Kentucky}\\
Dan Dale\\
Tim Gorringe\\
Wolfgang Korsch\\

\bigskip{\bf University of Lund}\\
Kevin Fissum\\

\bigskip{\bf University of Manitoba}\\
Juliette Mammei\\

\bigskip{\bf University of Maryland}\\
Elizabeth Beise\\

\bigskip{\bf University of Massachusetts, Amherst}\\
Krishna S. Kumar\\
Seamus Riordan\\
Jon Wexler\\

\bigskip{\bf University of New Hampshire}\\
Toby Badman\\
Trevor Bielarski\\
John Calarco\\
Bill Hersman\\
Maurik Holtrop\\
Donahy John\\
Mark   Leuschner\\
Elena Long\\
James Maxwell\\
Sarah Phillips\\
Karl Slifer\\
Timothy Smith\\
Ryan Zielinski\\

\bigskip{\bf University of Regina}\\
Garth Huber\\
George Lolos\\
Zisia Papandreou \\

\bigskip{\bf University of Saskatchewan}\\
Ru Igarashi\\

\bigskip{\bf University of Science and Technology of China (USTC)}\\
Yi Jiang\\
Wenbiao Yan \\
Yunxiu Ye\\
Zhengguo Zhao\\
Yuxian Zhao \\
Pengjia Zhu\\

\bigskip{\bf University of South Carolina}\\
Steffen Strauch\\

\bigskip{\bf University of Tennessee}\\
Nadia Fomin\\

\bigskip{\bf University of Virginia}\\
Khem Chirapatpimol\\
Donal Day\\
Xiaoyan Deng\\
Gordon D. Gates\\
Gu Chao\\
Charles Hanretty\\
Ge  Jin\\
Richard Lindgren\\
Jie Liu\\
Nilanga Liyanage\\
Vladimir Nelyubin\\
Blaine Norum\\
Kent Paschke\\
Peng Chao\\
Oscar Rondon\\
Kiadtisak Saenboonruang\\
William ``Al'' Tobias\\
Diancheng Wang\\
Kebin Wang\\
Zhihong Yi\\
Zhiwen Zhao\\
Xiaochao Zheng\\
Jiayao Zhou\\

\bigskip{\bf University of Washington}\\
Diana Parno\\

\bigskip{\bf Yamagata University}\\
Seigo Kato\\
Hiroaki Ueno\\

\bigskip{\bf Yerevan Physics Institute}\\
Sergey Abrahamyan\\
Nerses Gevorgyan\\
Edik Hovhannisyan\\
Armen Ketikyan\\
Samvel Mayilyan\\
Karen Ohanyan\\
Artush Petrosyan\\
Galust Sargsyan \\
Albert Shahinyan\\
Hakob Voskanian\\

\bigskip{\bf Former or Currently Inactive Members}\\
Mattias Anderson\\
Maud Baylac\\
Hachemi Benaoum\\
J. Berthot\\
Michel Bernard \\
Louis Bimbot\\
Tim Black\\
Alexander Borissov\\
Vincent Breton\\
Herbert Breuer\\
Etienne Burtin\\
Christian Cavata\\
George Chang\\
Nicholas Chant\\
Jean-Eric Ducret\\
Zhengwei Chai\\
Brandon Craver \\
Natalie Degrande\\
Rachel di Salvo\\
Pibero Djawotho\\
Chiranjib Dutta\\
Kim Egiyan\\
Stephanie Escoffier\\
Catherine Ferdi\\
Megan Friend \\
Robert Feuerbach\\
Mike Finn\\
Bernard Frois\\
Oliver Gayou\\
Charles Glashausser\\
Jackie Glister\\
Greg Hadcock\\
Brian Hahn\\
Harry Holmgren\\
Sebastian Incerti\\
Mauro Iodice\\
Riccardo Iommi\\
Florian Itard\\
Stephanie Jaminion\\
Steffen Jensen\\
Sudirukkuge Tharanga Jinasundera\\
Cathleen Jones\\
Lisa Kaufman\\
James D. Kellie\\
Sophie Kerhoas\\
Ameya Kolarkar\\
Norm Kolb\\
Ioannis Kominis\\
Serge Kox\\
Kevin Kramer\\
Elena Kuchina\\
Serguei Kuleshov\\
Jeff Lachniet\\
Geraud Lavessiere\\
Antonio Leone\\
David Lhuillier\\
Meihua Liang\\
Han Liu\\
Robert Lourie\\
Jacques Marroncle\\
Jacques Martino\\
Kathy McCormick\\
Justin McIntyre\\
Luis Mercado\\
Brian Milbrath\\
Wilson Miller\\
Joseph Mitchell\\
Jean  Mougey\\
Pierre Moussiegt\\
Alan Nathan\\
Damien Neyret\\
Stephane Platchkov\\
Thierry Pussieux\\
Gilles Quemener\\
Abdurahim Rakhman\\
Bodo Reitz\\
Rikki Roche\\
Philip Roos\\
David Rowntree\\
Gary Rutledge\\
Marat Rvachev\\
Arun Saha\\
Neil Thompson\\
Luminita Todor\\
Paul   Ulmer\\
Antonin Vacheret\\
Luc Van de Hoorebeke\\
Robert Van de Vyver\\
Pascal Vernin\\
Dan Watts\\
Krishni Wijesooriya\\
Hong Xiang\\
Wang Xu\\
Jingdong Yuan\\
Jianguo Zhao\\
Jingdong Zhou\\
Xiaofeng Zhu\\
Piotr Zolnierczuk\\
}

\end{multicols}

\end{document}